\documentclass[useAMS,usenatbib]{mn2e}

\usepackage[T1]{fontenc}
\usepackage{ae,aecompl}
\usepackage[dvips]{graphicx}
\usepackage{epsfig}
\usepackage{rotating}
\usepackage{graphicx}
\usepackage{caption}
\usepackage{amssymb}
\usepackage[caption=false]{subfig}
\usepackage{color}
\usepackage{hyperref}

\newcommand{\Msol}{\,{\rm M_{\odot}}}

\newcommand{\hMpc}{\, h^{-1}{\rm Mpc}}

\title[The accretion history of dark matter halos III]
{The accretion history of dark matter halos III: \\ A physical model for the concentration-mass relation}
\author[C.A.~Correa, J.S.B.~Wyithe, J.~Schaye and A.R.~Duffy]
  {Camila A.~Correa$^1$\thanks{E-mail: correac@student.unimelb.edu.au}, J. Stuart B.~Wyithe$^1$, Joop~Schaye$^2$ and Alan R.~Duffy$^{1,3}$\\
  $^1$ School of Physics, University of Melbourne, Parkville, VIC 3010, Australia\\
  $^2$ Leiden Observatory, Leiden University, PO Box 9513, 2300 RA Leiden, The Netherlands\\
  $^3$ Centre for Astrophysics and Supercomputing, Swinburne University of Technology, Melbourne, VIC 3122, Australia}
\date{\today}

\pagerange{\pageref{firstpage}--\pageref{lastpage}} \pubyear{2013}

\def\LaTeX{L\kern-.36em\raise.3ex\hbox{a}\kern-.15em
    T\kern-.1667em\lower.7ex\hbox{E}\kern-.125emX}

\begin{document}

\label{firstpage}

\maketitle

\begin{abstract}
We present a semi-analytic, physically motivated model for dark matter halo concentration as a function of halo mass and redshift. The semi-analytic model combines an analytic model for the halo mass accretion history (MAH), based on extended Press Schechter (EPS) theory, with an empirical relation between concentration and formation time obtained through fits to the results of numerical simulations. Because the semi-analytic model is based on EPS theory, it can be applied to wide ranges in mass, redshift and cosmology. The resulting concentration-mass ($c-M$) relations are found to agree with the simulations, and because the model applies only to relaxed halos, they do not exhibit the upturn at high masses or high redshifts found by some recent works. We predict a change of slope in the $z\sim 0$ $c-M$ relation at a mass scale of $10^{11}\Msol$. We find that this is due to the change in the functional form of the halo MAH, which goes from being dominated by an exponential (for high-mass halos) to a power-law (for low-mass halos). During the latter phase, the core radius remains approximately constant, and the concentration grows due to the drop of the background density. We also analyse how the $c-M$ relation predicted by this work affects the power produced by dark matter annihilation, finding that at $z = 0$ the power is two orders of magnitude lower than that obtained from extrapolating best-fitting $c-M$ relations. We provide fits to the $c-M$ relations as well as numerical routines to compute concentrations and MAHs\footnote{}. 
 \end{abstract}


\footnotetext{Available at \href{https://bitbucket.org/astroduff/commah}{\it{https://bitbucket.org/astroduff/commah}} and \href{http://astro.physics.unimelb.edu.au/Research/Public-Data-Releases/COMMAH}{\it{http://astro.physics.unimelb.edu.au/}} in Research/Public-Data-Releases/COMMAH} 

\section{Introduction}

Over the past few years large cosmological simulations have been performed to determine the properties of dark matter halos, including density profiles, shapes and accretion histories (see e.g. \citealt{Springel05,Klypin2011,Bryan13}). These properties are of particular interest, as forming galaxies depend on the structural properties of the halos in which they are embedded.

During hierarchical growth, halos acquire a density profile with a near universal shape, that can be described by a simple formula known as the `NFW profile' (\citealt{NFW97}, hereafter NFW). The NFW density profile is described by just two parameters, halo mass, $M$, and concentration, $c$. A halo's concentration is defined as the ratio of the virial radius, $R_{\rm{vir}}$, and the scale radius, $r_{-2}$, which is defined as the radius where the logarithmic density slope is $-2$. Thus, given the NFW profile, only a relation between concentration and halo mass (hereafter, the $c-M$ relation) is needed to fully specify halo structure at fixed mass. Therefore, numerous studies have been undertaken to improve the $c-M$ calibration.

Despite its importance, there is still no solid agreement on the dependence of halo concentration on halo mass and redshift. A small change in the adopted cosmology can have important effects on the structure of dark matter halos (\citealt{Maccio08}), and on their mass accretion histories (\citealt{Zhao09}). For example, the mean concentrations of dwarf-scale dark matter halos change by a factor of 1.5 between the various Wilkinson Microwave Anisotropy Probe (WMAP) cosmologies (\citealt{Spergel03,Spergel07}). The Planck cosmology (\citealt{Planck}) has higher matter density, $\Omega_{\rm{m}}$, and higher power spectrum normalization, $\sigma_{8}$, compared to the cosmological parameters of the year 5 data release of WMAP (WMAP5;~\citealt{Komatsu}). The Planck cosmology therefore suggests that halos assemble earlier and are more concentrated (c.f. $c-M$ relations from \citealt{Dutton14} and \citealt{Duffy08}). 

However, cosmology may not be the primary reason for the differences in the concentration$-$mass relations found by various authors. Recent works that adopt the same cosmology still find different $c-M$ relations (compare for example \citealt{Dutton14} and \citealt{Diemer14b}, or \citealt{Klypin2011} and \citealt{Prada}). \citet{Dutton14} found that the $c-M$ relation is well described by a power law, but flattens at high redshift and exhibits a positive slope at $z>4$. In contrast, \citet{Diemer14b} found a strong upturn in the high-mass end of the $c-M$ relation at all redshifts. The disparity between these studies could be due to the dynamical state of the selected dark matter halos. For example, \citet{Ludlow12} showed that massive halos that are substantially out of equilibrium are more likely to be found at a transient stage of high concentration, thus explaining the puzzling upturn in the high-mass end of the $c-M$ relation. Indeed, they reported that the upturn disappears when only dynamically-relaxed systems are considered. However, \citet{Klypin14} argued that the virial criterion used by \citet{Ludlow12} to select relaxed systems is incorrect, as it needs to include effects of the surface pressure and external forces. \citet{Klypin14} modified the virial criterion and ended up selecting massive halos that had previously been considered as unrelaxed. As a result, \citet{Klypin14} obtained an upturn in the $c-M$ relation of their relaxed halo sample and concluded that the upturn is a real feature of the $c-M$ relation. They explained that as extremely massive halos have more radial infall velocities, infalling mass penetrates deeper within the inner halo, thus increasing the concentration and producing the upturn. 

The main goal of this work is to derive a physically motivated model for the $c-M$ relation of relaxed halos based on the dark matter halo accretion history. By relating the concentration to the halo accretion history, we find that the $c-M$ relation does not show any upturn or strong flattening. We then study the $c-M$ relation in detail using simulations and selecting relaxed halos without using the virial criterion, and investigate whether recently accreted particles are able to reach the inner parts of the halo and thus increase the concentration. 

Our $c-M$ model relies on the fact that concentrations depend on the evolutionary stage of halos when they were formed. Several works have suggested that halo formation can be described as an `inside out' process, where a bound core (of a certain fraction of the halo mass today) collapses, followed by the gradual addition of material at the cosmological accretion rate (\citealt{Manrique,Wang09,Dalal,Salvador}). In this framework, the halo concentration should depend on the epoch at which a certain fraction of the halo mass was assembled. As a result, various authors (\citealt{Bullock,Wechsler,Zhao03,Ludlow14}) have provided models that relate $c$ to the halo mass history. For instance, \citet{Zhao03} showed that when the mass accretion rate of a halo slows down at low redshift, its scale radius, $r_{-2}$, remains approximately constant, and hence that concentration scales with the virial radius. On the other hand, in the regime of a high mass accretion rate (at high redshift), the scale radius scales approximately as the virial radius and thus $c$ remains constant.

The connection between a halo's mass accretion history (hereafter MAH) and its concentration, $c$, is therefore obtained through its `formation' time. The halo formation (or assembly) time is traditionally defined as the point in time when the halo mass reached a fraction of the total mass today. Low-mass halos typically assemble earlier, when the Universe was denser, than high-mass halos do. As a result, low-mass halos are more concentrated. Clearly, if concentration correlates with formation time, and formation time depends on the mass variance, $\sigma$ (because $\sigma$ describes the halo MAH, see the analytic model for the MAH from \citealt{PaperI}), then it is expected that $c$ correlates with $\sigma$ and hence with the peak height, $\nu$, defined as $\nu=1.686/\sigma$. This is indeed what several works have found (e.g. \citealt{Zhao09,Prada,Ludlow14,Dutton14}). We showed in \citet{PaperII} that the physical origin of the $c-\sigma$ (or $c-\nu$) relation is the halo MAH. 

Recently, in \citet{PaperI} and \citet{PaperII} (hereafter Paper I and Paper II, respectively), we provided two models for the MAH of halos, an analytic model and a semi-analytic model. The semi-analytic model uses a functional form for the MAH, that is motivated by extended Press-Schechter (EPS) theory, and links the MAH to halo structure through two empirical relations obtained from simulations. The analytic model is fully derived from the EPS formalism and thus does not require calibration against any simulation data. 

In the analytic model, the halo MAH is described in terms of the {\it{rms}} of the density perturbation field, $\sigma$, as ${M(z)=M_{0}(1+z)^{af(M_{0})}e^{-f(M_{0})z}}$, where $M_{0}$ refers to the present halo mass, $a$ depends on cosmology, and ${f(M_{0})\sim1/\sigma(M_{0})}$. This expression illustrates that as $\sigma$ decreases with halo mass, the function $f(M_{0})$ increases, causing the exponential in $M(z)$ to dominate. As a result, high-mass halos accrete faster than low-mass ones, due to their low value of $\sigma$. As low $\sigma$ implies large peak height,  the EPS formalism predicts that density perturbations with large $\nu$ experience an accelerated collapse phase relative to the average, and grow faster in time.

In this work we present a semi-analytic, physically motivated model for dark matter halo concentration as a function of halo mass, redshift and cosmology. The semi-analytic model, which builds on that of \citet{Ludlow14}, uses the analytic model for the halo MAH provided in Paper I, as well as an extension of the empirical relation between concentration and formation time obtained through fits to simulations provided in Paper II. As a result, the semi-analytic model for halo concentrations shows how the $c-M$ relation is expected to evolve based on the hierarchical growth of halos. 

This paper is organized as follows. We begin in Section~\ref{Simulations} with a description of the set of cosmological simulations used in this work. In Section~\ref{Halo_history_modelling}, we describe the analytic MAH model provided in Paper I and extend it to high redshift. In Section~\ref{cM_model}, we define halo formation time and build an empirical relation between formation time and concentration through fits to simulation data. Next, we describe the semi-analytic model for halo concentrations that combines the analytic model for the MAH and the empirical relation described previously. We analyse the evolution of concentration that predicts the semi-analytic model in Section~\ref{c_evolution}. In Section~\ref{DM_annihilation_signal} we discuss the impact of the results of our semi-analytic model for halo concentration on the signal from dark matter annihilation. In Section~\ref{Discussion} we discuss the main assumptions the semi-analytic model relies on. Finally, we summarize and conclude in Section~\ref{Conclusion}.

\section{Simulations}\label{Simulations}
Throughout this work we compare our analytic results to the output from numerical simulations. We use a set of cosmological dark matter only (DMONLY) simulations from the OWLS project (\citealt{Schaye}). These simulations were run with a significantly extended version of the N-Body Tree-PM, SPH code gadget3 (last described in \citealt{Springel05}). The initial conditions were generated with CMBFAST (version 4.1; \citealt{Seljak}) and evolved to redshift $z = 127$, where the simulations were started, using the \citet{ZelDovich} approximation from an initial glass-like state (\citealt{White96}). In order to assess the numerical convergence, we use simulations of different box sizes (ranging from 25 to 400 $\hMpc$) and particle numbers (ranging from $128^{3}$ to $512^{3}$). The simulation names contain strings of the form LxxxNyyy, where xxx is the simulation box size in comoving $\hMpc$ and yyy is the cube root of the number of particles. Our DMONLY simulations assume the WMAP5 cosmology. However, to investigate the dependence on the adopted cosmology, we use an extra set of five dark matter only simulations ($100 \hMpc$ box size and $512^{3}$ dark matter particles) which assume values for the cosmological parameters derived from different releases of the WMAP and the Planck missions. See the tables in Appendix \ref{Simulations_appendix} for the sets of cosmological parameters adopted in the different simulations, as well as the main numerical parameters of the runs such as comoving box size, number of dark matter particles, dark matter particle mass, comoving gravitational softening and maximum physical softening.

\section{Halo mass accretion history}\label{Halo_history_modelling}

We begin this section by briefly reviewing the analytic model for the MAH derived from the EPS formalism in Paper I, and showing how the MAH depends on cosmology and on the initial peak of the primordial density field. In Section \ref{MAHmodel_highz} we extend it to estimate the halo MAH tracked from an arbitrary redshift. Readers only interested in the $c-M$ relation model can skip directly to Section~\ref{cM_model}.

\subsection{Analytic model for the halo mass history}\label{MAH_model}

\begin{figure} 
\begin{center}
\includegraphics[angle=0,width=0.48\textwidth]{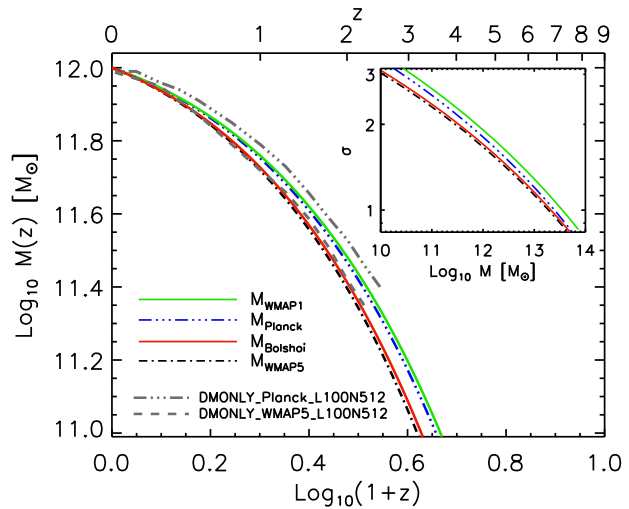}\\
\caption{Halo MAH of a $10^{12}\Msol$ halo (coloured lines) obtained from the model given by eqs. (\ref{MzLCDM_highz})-(\ref{zf_LCDM}), and by assuming various cosmologies as indicated in the legend. The grey lines correspond to MAH obtained from DMONLY simulations that assume the Planck and WMAP5 cosmologies. In the top right corner, we plot $\sigma$ versus halo mass, to show that the change in $\sigma$ under different cosmologies drives the change in the MAH.}
\label{Mz12_LCDM}
\end{center}
\end{figure}

In Paper I, we used simple analytic arguments based on the EPS formalism and the analytic formulation of \citet{Neistein}, to show that the `shape' of the MAH is determined by the growth factor of the initial density pertubations. The halo MAH is well described by an exponential in the high-redshift regime, but it slows to a power law at low redshift, because the growth of density perturbations is halted in the dark energy dominated era due to the accelerated expansion of the Universe. Therefore, we showed that the expression

\begin{equation}\label{MzLCDM}
M(z)_{\Lambda\rm{CDM}}=M_{0}(1+z)^{\alpha}e^{\beta z},
\end{equation}

\noindent accurately captures the median halo MAH, where $M_{0}$ refers to halo mass today, and $\alpha$ and $\beta$ are parameters that depend on $M_{0}$, cosmology and the linear power spectrum. In the case of an Einstein de Sitter (EdS) cosmology ($\Omega_{\Lambda}=0$ and $\Omega_{\rm{m}}=1$) or an open universe ($\Omega_{\Lambda}=0$ and $\Omega_{\rm{m}}<1$), there is no acceleration in the expansion of the Universe at low redshift. Then the halo mass history is simply described by an exponential as ${M(z)_{\rm{EdS}}=M_{0}e^{\beta z}}$, where $\beta=-1.686(2/\pi)^{1/2}f(M_{0})$. For a complete description of the model, see Paper I.

We find that the MAH model can be used to calculate halo mass histories in cosmologies other than WMAP5, and that the differences are mainly driven by the changes in $\sigma_{8}$ and $\Omega_{\rm{m}}$. We show this in Fig. \ref{Mz12_LCDM}, where the halo MAH of a $10^{12}\Msol$ halo (coloured lines) was estimated for the various cosmologies, as indicated in the legend. In the top right corner of Fig. \ref{Mz12_LCDM}, we plot $\sigma$ versus halo mass, to show how the change in $\sigma$ drives the change in the MAH. The exception is the Planck cosmology, which has a relatively low $\sigma_{8}$ but a large $\Omega_{\rm{m}}=0.317$, which raises $M(z)$ close to the WMAP1 $M(z)$.

The overplotted grey lines in Fig.~\ref{Mz12_LCDM} correspond to the MAH obtained from DMONLY simulations that assume the Planck and WMAP5 cosmologies. In this case, we compute the MAH of the main subhalo (that is not embedded inside a larger halo) of Friends-of-Friends (FoF) groups (\citealt{Davis}), by tracking the virial mass of the main progenitor at each prior output redshift. Halo virial masses and radii were determined using a spherical overdensity routine within the SUBFIND algorithm (\citealt{Springel}) centred on the main subhalo of FoF halos. Throughout this work we define the halo mass as the total mass within the radius $r_{200}$ for which the mean internal density is 200 times the critical density. For a more detailed description of the method used to create merger trees, see Paper II.

\begin{figure*} 
\begin{center}
\includegraphics[angle=0,width=0.8\textwidth]{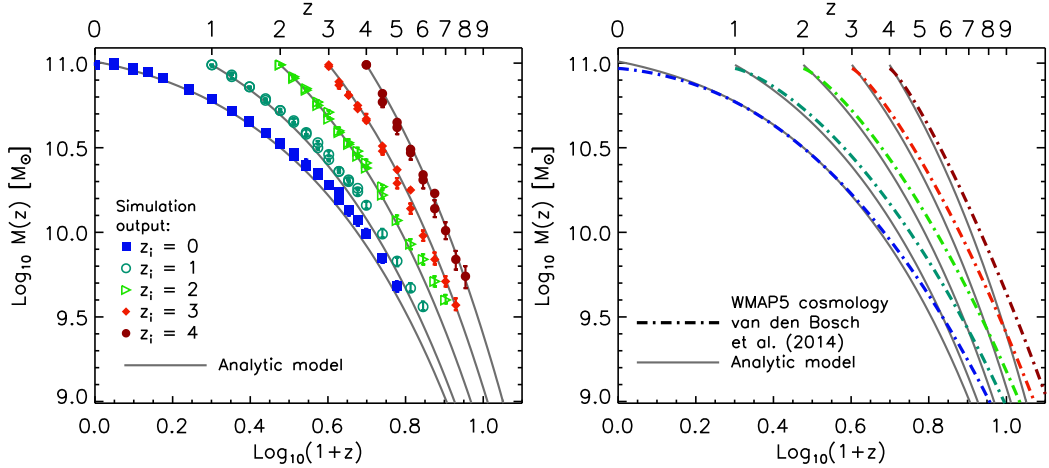}\\
\caption{Median MAHs for halos of $10^{11}\Msol$  starting from various redshifts. In both panels the grey solid lines correspond to the analytic model described in Section~\ref{MAHmodel_highz}. The coloured curves in the left panel are the MAHs obtained from the DMONLY simulations WMAP5$_{-}$L025N512 and WMAP5$_{-}$L050N512. The mass histories are computed by calculating the median value and the $1\sigma$ error bars are determined by bootstrap resampling the halos from the merger tree at a given output redshift. The coloured dot dashed curves in the right panel are the MAHs obtained from the \citet{vandenBosch14} model.}
\label{MAH_z0}
\end{center}
\end{figure*}

\subsection{Analytic model for the MAH: high redshift prediction}\label{MAHmodel_highz}

The model presented in Paper I is suitable for estimating halo MAHs that are tracked from ${z=0}$. In this section we extend this analytic model to estimate MAHs of halos of the same halo mass that are tracked from arbitrary redshifts $z_{i}$. This is shown in Fig. \ref{MAH_z0}, where the MAHs of $10^{11}\Msol$ halos are obtained from DMONLY simulations (coloured curves). The curves show the mean MAH of halos of the same mass ($10^{11}\Msol$ in this case) that begin at $z_{i}=0$ (blue curve), $1$ (dark green curve), $2$ (green curve), $3$ (orange curve) and $4$ (red curve). High-redshift MAHs are dominated by large accretion rates and characterized by a pure exponential. 

We generalize the analytic model so that it describes the MAHs from any $z_{i}$ redshift. Expression (\ref{MzLCDM}) can be rewritten as 

\begin{equation}\label{MzLCDM_highz}
\tilde{M}(z,M(z_{i}),z_{i})= M(z_{i})(1+z-z_{i})^{\tilde{\alpha}}e^{\tilde{\beta}(z-z_{i})},
\end{equation}

\noindent where $\tilde{M}(z,z_{i})$ denotes the MAH of a halo with mass $M(z_{i})$ at redshift $z_{i}$. In the above expression, $z>z_{i}$ and the parameters $\tilde{\alpha}$ and $\tilde{\beta}$ depend on ${M(z_{i})}$ and redshift $z_{i}$

\begin{eqnarray}\label{alpha_tilde}
\tilde{\alpha} &=& \left[\frac{1.686(2/\pi)^{1/2}}{D(z_{i})^{2}}\frac{dD}{dz}|_{z=z_{i}}+1\right]f(M(z_{i})),\\\label{beta_tilde}
\tilde{\beta} &=& -f(M(z_{i})),\\\label{fM}
f(M(z_{i})) &=& [\sigma^{2}(M(z_{i})/q)-\sigma^{2}(M(z_{i}))]^{-1/2},\\\label{sigma_def}
\sigma^{2}(R) &=& \frac{1}{2\pi^{2}}\int_{0}^{\infty}P(k)\hat{W}^{2}(k;R)k^{2}dk,\\\label{q_LCDM}
q&=&4.137\times z_{\rm{f}}^{-0.9476},\\\label{zf_LCDM}
z_{\rm{f}}&=&-0.0064(\log_{10}M_{0})^{2}+0.0237(\log_{10}M_{0})\\\nonumber
& & +1.8837,
\end{eqnarray}

\noindent where $D(z)$ is the linear growth factor, $P(k)$ the linear power spectrum, $\hat{W}(k;R)$ the Fourier transform of a top hat window function and $R$ defines $\sigma$ in a sphere of mass $M=(4\pi/3)\rho_{\rm{m},0}R^{3}$, where $\rho_{\rm{m},0}$ is the mean background density today. We use the approximation of \citet{Eisenstein} to compute $P(k)$, normalized such that $\sigma(8h^{-1}\rm{Mpc})=\sigma_{8}$.  As a result, $f(M_{0})$ depends on the power spectrum and halo mass. It can be seen from eqs.~(\ref{alpha_tilde}) and (\ref{beta_tilde}) that at large $z_{i}$, $\tilde{\alpha}\rightarrow 0$ due to $D(z_{i})\propto 1.686(2/\pi)^{1/2}/(1+z_{i})$ for $z_{i}\gg 1$, indicating that the MAH is mainly described by an exponential. Table \ref{Notation} provides a summary of the nomenclature adopted throughout this work.
The above equations introduce an analytic halo MAH model directly derived from EPS theory that does not require calibration against any simulation data (see Paper I for more details). The numerical values given in eqs. (\ref{q_LCDM}) and (\ref{zf_LCDM}) were determined by assuming the WMAP5 cosmology (${\Omega_{\rm{m}}=0.258}, {\Omega_{\Lambda}=0.742}, {h=0.72}, {n_{s}=0.963}, {\sigma_{8}=0.796}$).
 
In the left panel of Fig.~\ref{MAH_z0}, we compare the model given by eqs.~(\ref{alpha_tilde})-(\ref{zf_LCDM}) to various MAHs obtained from a set of DMONLY simulations. Our analytic model is shown by grey solid lines, where we have taken ${M(z_{i})=10^{11}\Msol}$. The coloured curves in the left panel correspond to the MAHs obtained from the DMONLY simulations WMAP5$_{-}$L025N512 and WMAP5$_{-}$L050N512. We find very good agreement between the simulation outputs and the analytic model at all redshifts. The simulation outputs from the boxes $L=25\hMpc$ and $L=50\hMpc$ converge up to $z=5$. At higher $z$, the outputs from the $L=25\hMpc$ simulation underestimate $M(z)$ because the box size limits the maximum sizes of the structures that can form at each redshift.

In the right panel of Fig.~\ref{MAH_z0}, we compare our extended analytic model with the \citet{vandenBosch14} model. \citet{vandenBosch14} extracted halo mass histories from the Bolshoi simulations (\citealt{Klypin2011}) and extended them below the numerical resolution limit using EPS merger trees. Once they had obtained the MAH curves for a large range of redshifts and halo masses, they made use of a semi-analytic model to transform the (average or median) MAHs, based on the Bolshoi cosmology, to other cosmologies. Using their publicly available code, we calculate the mass history curves for the WMAP5 cosmology for comparison with our results. We find that there is some discrepancy at high-redshift for all the curves. The \citet{vandenBosch14} MAH model seems to over predict the halo mass at $z>5$, most likely as a consequence of the different halo definitions, and subtle differences in the definition of the main progenitor (van den Bosch, private communication). Overall, there is very good agreement between the most recent accretion history study in the literature and our model, as well as with the simulation outputs. In Section~\ref{cM_model} we will make use of our analytic MAH model to calculate concentrations.

Using the extended MAH model for high redshift, we can calculate the accretion rate of a halo at redshift $z$. We differentiate eq.~(\ref{MzLCDM_highz}) with respect to time and replace $dz/dt$ by $-H_{0}[\Omega_{\rm{m}}(1+z)^{5}+\Omega_{\Lambda}(1+z)^{2}]^{1/2}$, to obtain

\begin{eqnarray}\nonumber
\frac{d\tilde{M}(z,M(z_{i}),z_{i})}{dt} &=& 71.6{\rm{M}}_{\sun}{\rm{yr}}^{-1} \left(\frac{\tilde{M}(z,M(z_{i}),z_{i})}{10^{12}\rm{M}_{\sun}}\right) \\\nonumber
&\times & \left(\frac{h}{0.7}\right)[-\tilde{\alpha}/(1+z-z{i})-\tilde{\beta}]\\\label{dMdt}
&\times &  (1+z)[\Omega_{\rm{m}}(1+z)^{3}+\Omega_{\Lambda}]^{1/2},
\end{eqnarray}

\noindent where $\tilde{\alpha}$ and $\tilde{\beta}$ are given by eqs.~(\ref{alpha_tilde}) and (\ref{beta_tilde}), respectively. Note that the above formula will give the accretion rate at redshift $z$ of a halo that has mass $M(z_{i})$ at redshift $z_{i}$, and mass $\tilde{M}(z,M(z_{i}),z_{i})$ at redshift $z$.

\section{Concentration$-$ mass relation}\label{cM_model}

A theoretical understanding of the physical connection between concentration (the parameter that characterizes the internal structure of NFW dark matter halos) and the initial conditions of the density field, is essential for the physical interpretation of relations like $c-\nu$ (concentration$-$peak height) or $c-M$, that have been calibrated using cosmological simulations (e.g. \citealt{Bullock,Neto,Gao,Maccio07,Duffy08,Ludlow13,Dutton14,Diemer14b}). 

It has previously been shown that concentration is determined by the halo MAH, and that the MAH depends on the power spectrum and the adopted cosmological parameters (\citealt{Wechsler,Zhao03,Ludlow13,Ludlow14}). In this section we show, through analytic and numerical modelling, how the concentration of dark matter halos depends on cosmology and the power spectrum of density perturbations. Our results imply that the halo MAH is the physical link between concentration and peak height.

\begin{figure*} 
  \begin{center}
  \includegraphics[angle=0,width=0.75\textwidth]{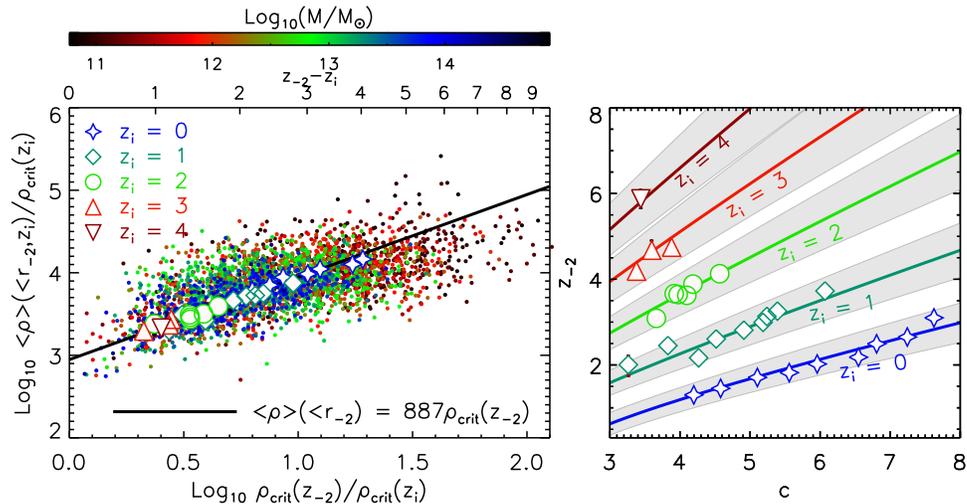}
  \end{center}
\caption{{\it{Left panel:}} mean density within the NFW scaled radius, $\langle\rho\rangle(<r_{-2})$, at $z_{i}$ against the critical density of the universe at the formation time, $\rho_{\rm{crit}}(z_{-2})$. Each dot in the panel corresponds to an individual relaxed halo identified at $z_{i}$ and coloured by mass according to the colour bar at the top of the plot. The open symbols show the median values of the sample at $z_{i}$ as indicated by the legend in logarithmic mass bins of width $\delta\log_{10}M=0.4$. The solid line shows the best linear fit to the  $\rho_{\rm{crit}}(z_{-2})-\langle\rho\rangle(<r_{-2})$ relation. {\it{Right panel:}} formation redshift against concentration. The solid lines show the $c-z_{-2}$ relations given by equation (\ref{zf_relation}) for various $z_{i}$ as indicated in the legend of the left panel. The open symbols correspond to the median values of the samples in logarithmic mass bins of width $\delta\log_{10}M=0.4$ and are colour coded by $z_{i}$. The grey areas show the scatter in $z_{-2}$.}
\label{formation_time}
\end{figure*}

\subsection{Formation redshift}\label{zf_definition}

As discussed in the Introduction, halo MAHs can be used to estimate halo concentrations. Halo concentrations reflect the mean density of the Universe at the formation redshift (NFW;\citealt{Wechsler,Zhao03,Zhao09,Ludlow13}). Therefore, the essential link between a halo's MAH and its internal structure is the formation redshift. For a halo with mass $M(z_{i})$ at redshift $z_{i}$, we define the formation redshift to be $z_{-2}$, the redshift at which the mass of the main progenitor equals the mass enclosed within the scale radius at $z=z_{i}$, 
\begin{equation}\label{z2}
z_{-2}=z[\tilde{M}(z_{-2},M(z_{i}),z_{i})=M_{r}(r_{-2},z_{i})],
\end{equation}

\noindent (\citealt{Ludlow13}). Here $\tilde{M}(z_{-2},M(z_{i}),z_{i})$ is the mass at $z_{-2}$ of a halo with mass $M(z_{i})$ at $z_{i}$, and we denote the mass enclosed within $r$, $M(<r)$, as $M_{r}$. For an NFW profile the internal mass $M_{r}(r_{-2},z_{i})$ is related to the total halo mass as 

\begin{equation}\label{M2}
M_{r}(r_{-2},z_{i})=M(z_{i})\frac{Y(1)}{Y(c[M(z_{i}),z_{i}])},
\end{equation}

\noindent where $Y(u)=\ln(1+u)-u/(1+u)$, $c[M(z_{i}),z_{i}]$ is the concentration at $z_{i}$ and $M(z_{i})$ is the total halo mass at $z_{i}$. In cases where we identify halos at $z_{i}=0$ and track their mass histories, we calculate $z_{-2}$ by setting $M(z_{-2})$ equal to the mass enclosed within $r_{-2}$ today. In cases where we identify halos at $z_{i}>0$, we first calculate $r_{-2}$ and $M_{r}(r_{-2},z_{i})$ at the particular redshift $z_{i}$, and then find $z_{-2}$ by tracking the MAH (for $z>z_{i}$) and equating $\tilde{M}(z_{-2},M(z_{i}),z_{i})$ to $M_{r}(r_{-2},z_{i})$. See Table~\ref{Notation} for a summary of the nomenclature.

\citet{Ludlow13,Ludlow14} and Paper II showed that $z_{-2}$ correlates strongly with $c$, and in Paper II we demonstrated that the scatter in $z_{-2}$ and in the halo MAH predicts the scatter in $c$. In this section we explore how the formation time $-$ concentration relation varies for halos identified at various redshifts.

We computed density profiles and MAHs for halos identified at redshifts $z_{i} = 0, 1, 2, 3$ and $4$. The density profiles were computed by fitting the NFW density profile,

\begin{equation}\label{NFW}
\rho(r,z_{i})=\frac{\rho_{\rm{crit}}(z_{i})\delta_{\rm{c}}}{(cr/r_{200})[1+(cr/r_{200})]^{2}},
\end{equation}

\noindent for each individual halo. In the above equation, ${\rho_{\rm{crit}}(z_{i})=3H^{2}(z_{i})/8\pi G}$ is the critical density of the universe, $\delta_{\rm{c}}$ is a dimensionless parameter related to the concentration $c=r_{200}/r_{-2}$ by $\delta_{\rm{c}}=\frac{200}{3}\frac{c^{3}}{Y(c)}$ and $r_{200}$ is the virial radius. 

\begin{table}
\centering  
\caption{Notation reference.} 
\label{Notation}
\begin{tabular}{ll}
\hline
  Notation & Definition \\ \hline\hline
$M(z_{i})$ & Total halo mass at $z_{i}$,\\
& defined as $M_{200}$\\
$r_{200}$ or $r_{200}[M(z_{i}),z_{i}]$ & Virial radius at $z_{i}$ of  a halo of\\
& total mass $M(z_{i})$\\
$r_{-2}$ or $r_{-2}[M(z_{i}),z_{i}]$ & NFW scale radius at $z_{i}$\\
$c$ or $c[M(z_{i}),z_{i}]$ & NFW concentration at $z_{i}$\\
$M_{r}(r,z_{i})$ & $M(<r)$, mass enclosed within $r$\\
& at $z_{i}$ of a halo of total mass $M(z_{i})$\\
$M_{r}(r_{-2},z_{i})$ & Mass enclosed within $r_{-2}$ at $z_{i}$\\
$\tilde{M}(z,M(z_{i}),z_{i})$ & Mass at $z$ of a halo with mass\\
& $M(z_{i})$ at $z_{i}$\\
$z_{-2}$ & Formation redshift, when equating\\ 
& $\tilde{M}(z_{-2},M(z_{i}),z_{i})$ to $M_{r}(r_{-2},z_{i})$\\
$\langle\rho\rangle(<r_{-2},z_{i})$ & Mean density within $r_{-2}$ at $z_{i}$\\
$\rho_{\rm{crit},0}$ & Critical density today\\
$\rho_{\rm{crit}}(z_{i})$ & Critical density at $z_{i}$\\
\hline
\end{tabular}
\end{table}

\begin{figure*} 
  \begin{center}
  \includegraphics[angle=0,width=0.95\textwidth]{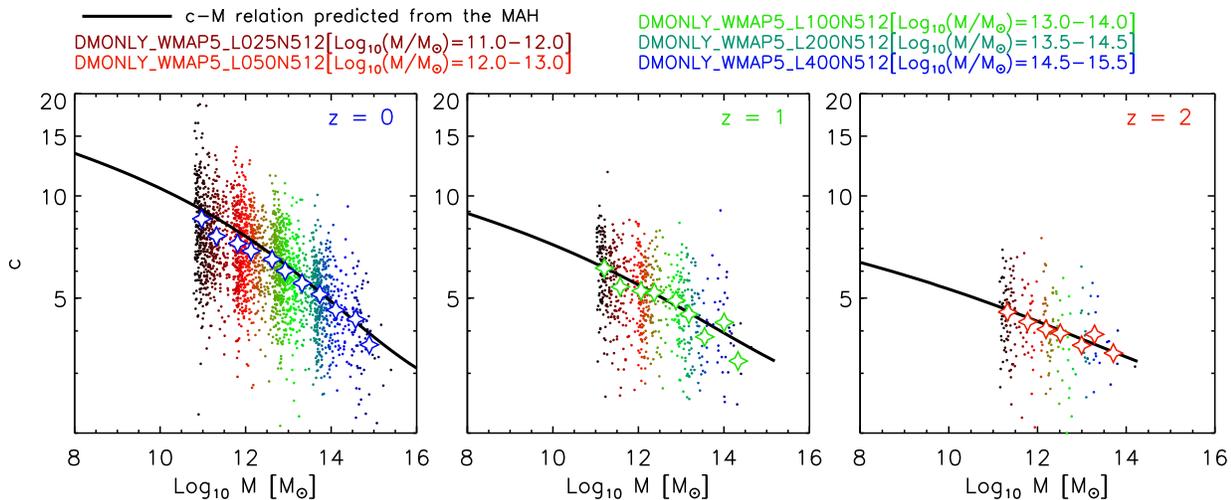}
  \end{center}
\caption{Concentration$-$mass relations at $z=0$ (left panel), $z=1$ (middle panel) and $z=2$ (right panel). The dots in the panels correspond to individual, relaxed halos illustrating the scatter in the relation. The simulations assume the WMAP5 cosmological parameters and have box sizes of 400, 200, 100, 50 and 25 $\hMpc$, as indicated. Because of resolution limits only halos in the mass ranges indicated in the top legend were used from a particular simulation. The open symbols show the median $c-M$ values in logarithmic mass bins of width $\delta\log_{10}M=0.4$. The solid line shows the prediction of the $c-M$ model obtained from the halo MAH as described in Section~\ref{cM_model}.}
\label{3plots}
\end{figure*}

We begin by fitting NFW profiles to all halos at $z_{i}$ that contain at least $10^{4}$ dark matter particles within the virial radius. Throughout this work, we define the virial radius as $r_{200}$, the radius for which the mean internal density is 200 times the critical density. Then, for each halo, all particles in the range $-1.25\le \log_{10}(r/r_{200})\le 0$ are binned radially in equally spaced logarithmic bins of size $\Delta\log_{10}r=-0.078$. The density profile is then fitted to these bins by performing a least-squares minimization of the difference between the logarithmic densities of the model and the data, assuming equal weighting. The corresponding mean enclosed mass, $M_{r}(r_{-2},z_{i})$, and mean inner density at $r_{-2}$, $\langle\rho\rangle(<r_{-2},z_{i})$, are found by interpolating along the cumulative mass and density profiles from $r=0$ to $r_{-2}=r_{200}/c$, where $c$ is the concentration from the fit of the NFW halo. Then we generate merger trees for these halos and by interpolation we determine the redshift $z_{-2}$ at which $\tilde{M}(z_{-2},z_{i})=M_{r}(r_{-2},z_{i})$. 

In order to obtain robust estimates and to test whether the $c-M$ relation includes an upturn in the median concentrations of massive halos (\citealt{Prada,Dutton14,Diemer14b}), we only consider `relaxed' halos. We define relaxed halos as those halos for which the separation between the most bound particle and the centre of mass of the FoF halos is smaller than $0.07R_{\rm{vir}}$ (following \citealt{Maccio07}, \citealt{Neto} and \citealt{Duffy08}), where $R_{\rm{vir}}$ is the radius within which the mean density is $\Delta$, as given by \citealt{BryanNorman}, times the critical density. Our relaxed sample contains 2425 halos at $z=0$, 726 halos at $z=1$, 226 halos at $z=2$ and 78 and 20 halos at $z=3$ and $z=4$, respectively. 

The left panel of Fig. \ref{formation_time} shows the mean density within the NFW scale radius, $r_{-2}$, at redshift $z_{i}$. The median values of $\langle\rho\rangle(<r_{-2},z_{i})$ follow the best-fitting relation

\begin{equation}\label{rho2}
\langle\rho\rangle(<r_{-2},z_{i})=200\frac{c[M(z_{i}),z_{i}]^{3}Y(1)}{Y(c[M(z_{i}),z_{i}])}\rho_{\rm{crit}}(z_{i}),
\end{equation}

\noindent expressed as a function of the critical density of the Universe at $z_{-2}$,

\begin{equation}\label{rho_crit}
\rho_{\rm{crit}}(z_{-2})=\rho_{\rm{crit,0}}[\Omega_{\rm{m}}(1+z_{-2})^{3}+\Omega_{\Lambda}],
\end{equation}

\noindent where $\rho_{\rm{crit,0}}=3H^{2}(z=0)/8\pi G$. Note that densities along both the $x-$ and $y-$axes are expressed in units of the critical density at $z_{i}$. Each dot in the panel corresponds to an individual relaxed halo identified at $z_{i}$ and coloured by mass according to the colour bar at the top of the plot. The open symbols show the median value of the sample in logarithmic mass bins of width $\delta\log_{10}M=0.4$ and are coloured by $z_{i}$ as indicated in the legend\footnote{Note that it is possible for individual halos to appear multiple times in Fig.~\ref{formation_time} (left panel). For example, a $10^{13}\Msol$ halo at $z=0$, has a total mass of $\sim 10^{12.2}\Msol$ at $z=2$, therefore the halo will be included in the $\rho_{\rm{crit}}(z_{-2})-\langle\rho\rangle(<r_{-2},z_{i})$ relation at $z_{i}=0$ but also at $z_{i}=2$.}.

At each redshift $z_{i}$, the $\rho_{\rm{crit}}(z_{-2})-\langle\rho\rangle(<r_{-2},z_{i})$ correlation clearly shows that halos which collapsed earlier have denser cores. 

We perform a least-squares minimization of the quantity $\Delta^{2} = \frac{1}{N}\sum_{j=1}^{N}[\langle\rho_{j}\rangle(<r_{-2},z_{i})-F(\rho_{\rm{crit},j}(z_{-2}),A)]$, where $j$ goes from 1 to the number of dark matter halos, N, at $z_{i}$ and $F(\rho_{\rm{crit},j}(z_{-2}),A)=A\times \rho_{\rm{crit},j}(z_{-2})$, to obtain the constant of proportionality, $A$. The solid line corresponds to the best-fit to the $\rho_{\rm{crit}}(z_{-2})-\langle\rho\rangle(<r_{-2},z_{i})$ relation, and we find (in agreement with Ludlow et al. 2014) that the average relation

\begin{equation}\label{rho_formation}
\frac{\langle\rho\rangle(<r_{-2},z_{i})}{\rho_{\rm{crit}}(z_{i})} = A\times \frac{\rho_{\rm{crit}}(z_{-2})}{\rho_{\rm{crit}}(z_{i})}
\end{equation}

\noindent is maintained through time with $A=887\pm 36$, where the $1\sigma$ error was obtained from the least-squares fit.

Using eqs. (\ref{rho2}) and (\ref{rho_crit}) we can rewrite this relation as 

\begin{equation}\label{zf_relation}
\frac{c[M(z_{i}),z_{i}]^{3}Y(1)}{Y(c[M(z_{i}),z_{i}])} = \frac{A}{200}\frac{[\Omega_{\rm{m}}(1+z_{-2})^{3}+\Omega_{\Lambda}]}{[\Omega_{\rm{m}}(1+z_{i})^{3}+\Omega_{\Lambda}]},
\end{equation}

\noindent The right panel of Fig. \ref{formation_time} shows the ${c-z_{-2}}$ relation (solid lines) given by  eq.~(\ref{zf_relation}) for various $z_{i}$. The open symbols correspond to the median values of the sample in logarithmic mass bins of width $\delta\log_{10}M=0.4$. The grey areas show the scatter in $z_{-2}$.

\subsection{Semi-analytic model for halo concentration}\label{halo_concentration}

In this section we describe the semi-analytic model for halo concentration as a function of halo mass and redshift. This model combines the analytic model for the halo MAH given by eqs. (\ref{MzLCDM_highz}-\ref{zf_LCDM}) and the empirical relation between $z_{-2}$ and $c$ given by eq. (\ref{zf_relation}). 

We begin by calculating $\tilde{M}(z_{-2},M(z_{i}),z_{i})$ from eq.~(\ref{MzLCDM_highz}), and use the equality

\begin{equation}
\frac{\tilde{M}(z_{-2},M(z_{i}),z_{i})}{M(z_{i})}=\frac{M_{r}(r_{-2},z_{i})}{M(z_{i})}=\frac{Y(1)}{Y(c[M(z_{i}),z_{i}])},
\end{equation}

\noindent which follows from eqs. (\ref{z2}) and (\ref{M2}) and is valid under the assumption that the halo density profile follows the NFW profile, to obtain

\begin{equation}\label{eq_c1}
\frac{Y(1)}{Y(c[M(z_{i}),z_{i}])}=(1+z_{-2}-z_{i})^{\tilde{\alpha}}e^{\tilde{\beta}(z_{-2}-z_{i})},
\end{equation}
 
\noindent where $\tilde{\alpha}$ and $\tilde{\beta}$ are given by eqs. (\ref{alpha_tilde}) and (\ref{beta_tilde}), respectively. Next, we combine eqs.~(\ref{zf_relation}) and (\ref{eq_c1}) to obtain the concentration, $c[M(z_{i}),z_{i}]$, of a halo of total mass $M(z_{i})$ at $z_{i}$. We remind the reader that throughout this work the adopted halo mass definition is $M_{200}$, and the concentrations are therefore defined as $c=c_{200}$.

Fig. \ref{3plots} shows the concentration$-$mass relation at $z=0$ (left panel), at $z=1$ (middle panel), and at $z=2$ (right panel). The dots in the panels correspond to individual relaxed halos identified in the simulations at $z_{i}=0,1$ and $2$, whereas the open symbols correspond to the median values in logarithmic mass bins of width $\delta\log_{10}M=0.4$. The solid line shows the $c-M$ relation that results from the semi-analytic model described above. We find excellent agreement between the median values from the simulations and the $c-M$ relation predicted by the semi-analytic model at all redshifts.

So far we have adopted the WMAP5 cosmology. In Appendix \ref{cosmo_dependence}, we discuss the dependence of our $c-M$ relation model on cosmology and extend it to make it suitable for any values of the cosmological parameters. 

\begin{figure*} 
  \begin{center}
  \includegraphics[angle=0,width=\textwidth]{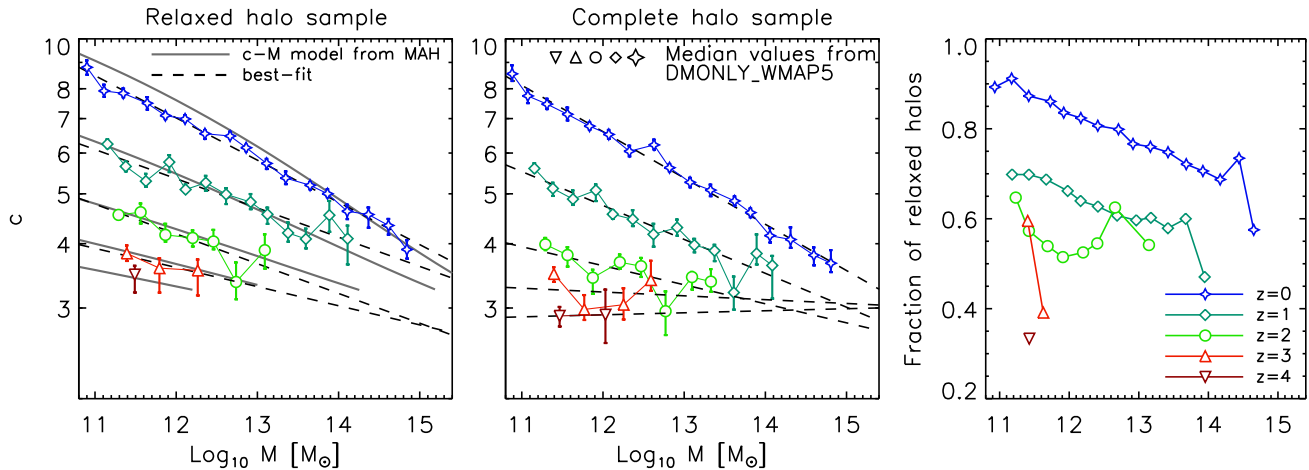}
  \end{center}
  \caption{Concentration $-$ mass relation at $z=0, 1, 2, 3$ and $4$ under the WMAP5 cosmology for the relaxed halo sample (left panel) and the full sample (middle panel). The open symbols indicate the median concentrations in logarithmic mass bins of width $\delta\log_{10}M=0.25$ at $z=0$ and $z=1$, and $\delta\log_{10}M=0.30$, $0.40$ and $0.50$ at $z=2$, $3$ and $4$, respectively. Only bins containing at least 10 halos are shown. The error bars show the $1\sigma$ confidence limits. The dashed lines correspond to the best-fitting power laws to the open symbols. In the left panel, the solid grey line shows the $c-M$ relation predicted by the semi-analytic model. The right panel shows the fraction of relaxed halos, with respect to the complete sample, for each mass bin and redshift. The inclusion of unrelaxed halos results in a flattening of, or even an upturn in, the $c-M$ relation at high redshift.}
\label{relaxed_comparison}
\end{figure*}

\subsection{Impact of relaxedness on the $c-M$ relation}

Several recent studies (\citealt{Klypin2011,Prada,Dutton14,Diemer14b}) have found that the $c-M$ relation flattens at high redshift and exhibits an `upturn' at the high-mass end, meaning that the concentration increases with halo mass for the most massive halos. In this section we investigate whether this interesting behaviour is seen in our semi-analytic model or in the simulation outputs. 

Our model does not predict an upturn. The model relates $c$ to the MAH via the formation redshift, $z_{-2}$ (see Fig. \ref{formation_time}, right panel), which decreases with halo mass, meaning that more massive halos are less concentrated because they formed more recently. If $c$ were to increase with halo mass, then high-mass halos would have to form earlier than low-mass ones, at a point when the Universe was denser. This behaviour is neither seen in our simulations (see Fig. \ref{Mz12_LCDM}, coloured lines), as we only consider relaxed systems, nor predicted by EPS theory, because it would be antihierarchical for such systems. 


\begin{figure*} 
  \begin{center}
  \includegraphics[angle=0,width=\textwidth]{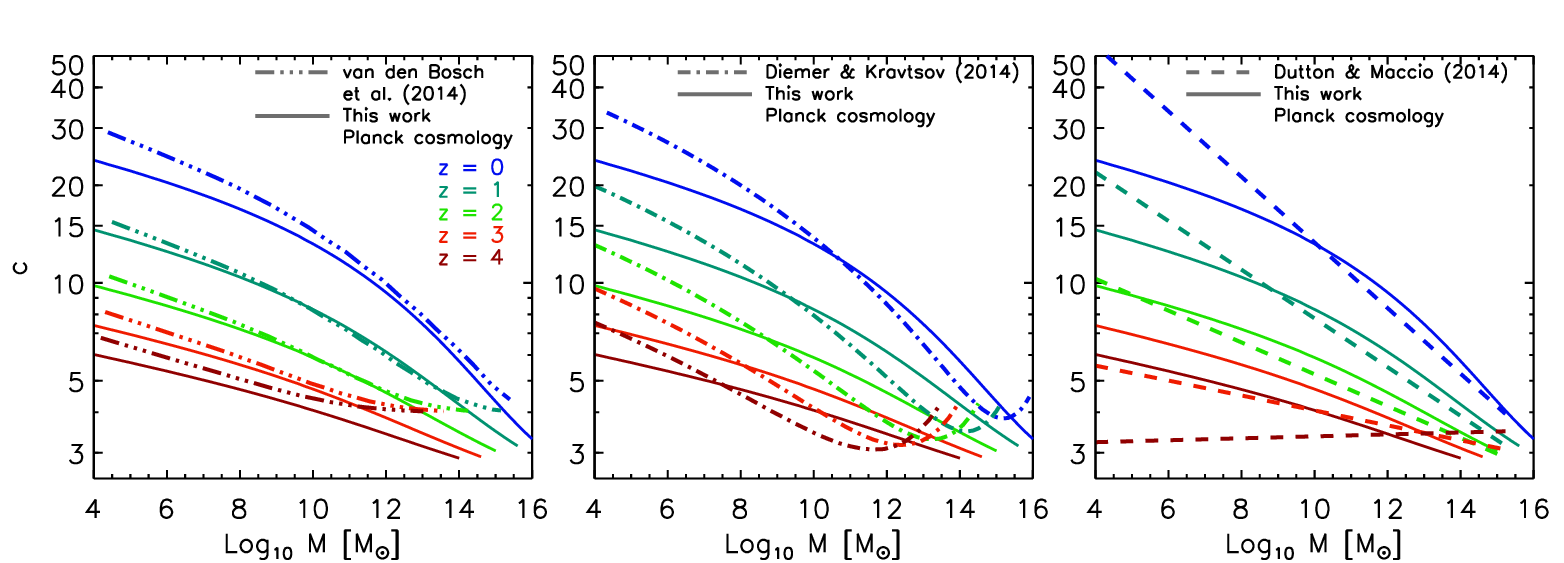}
  \end{center}
\caption{Comparison of the $c-M$ relation predicted by our model (solid lines) with the relations of \citet{vandenBosch14} (left-panel), \citet{Diemer14b} (middle-panel) and \citet{Dutton14} (right-panel). Note that \citet{Dutton14} consider only relaxed halos for their best-fitting model and fit their model in the halo mass range $10^{10}-10^{15}\Msol$, whereas \citet{Diemer14b} and \citet{vandenBosch14} use their full halo sample and fit their model in halo mass range $10^{10}-10^{15}\Msol$ and $10^{11}-10^{15}\Msol$, respectively.} 
\label{comparison}
\end{figure*}

To investigate further, we use the simulation outputs to calculate concentrations by fitting NFW profiles to halos that are resolved with at least $10^{4}$ particles within the virial radius, and for which the convergence radius\footnote{The convergence radius is defined such that the two$-$body dynamical relaxation time-scale of the particles is similar to the age of the universe. For more details see \citet{Power03} or \citet{Duffy08}.} (\citealt{Power03}) is smaller than the minimum fit radius of 0.05 times the virial radius. In addition, we consider two halo samples. A relaxed halo sample\footnote{As proposed by \citet{Neto}, relaxed halos are defined as those halos for which the separation between the most bound particle and the centre of mass of the FoF halo is smaller than 0.07 times the virial radius.} and a full halo sample. When considering only relaxed halos, as we have done so far, we find that we restrict our halo sample to around $80\%$ of the total at $z=0$, $65\%$ at $z=1$, $55\%$ at $z=2$, $50\%$ at $z=3$ and $43\%$ at $z=4$.

Fig.~\ref{relaxed_comparison} shows the $c-M$ relation (at various $z$) of the relaxed sample (left panel) and the full sample (middle panel). These panels show the median value of the concentration (open symbols) in logarithmic mass bins of width $\delta\log_{10}M=0.25$ at $z=0$ and $1$, and $\delta\log_{10}M=0.30$, $0.40$ and $0.50$ at $z=2$, $3$ and $4$, respectively. We increase the bin size with redshift so that each bin at a fixed mass contains on average approximately the same number of halos. For each bin the $1\sigma$ error bars were determined by bootstrap resampling the halos. Only bins containing at least 10 halos are shown. The dashed lines correspond to the best-fitting power laws to the open symbols. In addition, the left panel shows the $c-M$ relations predicted by the semi-analytic model in the solid grey lines. The middle panel shows a strong flattening and upturn in the $c-M$ relation at high $z$, in agreement with \citet{Munoz} and \citet{Prada}. However, this upturn is not seen for the relaxed sample. Thus, we conclude that the previously seen upturn results from the inclusion of unrelaxed halos, in agreement with \citet{Ludlow12}. We show the fraction of relaxed halos (with respect to the total sample) for each mass bin and redshift in the right panel of Fig.~\ref{relaxed_comparison}. We find that the relaxed fraction tends to decrease towards high mass and redshift.
 
Our results suggest that the dynamical state of dark matter halos should be considered when analysing the parameters that describe the halo internal structure, because the density profiles of unrelaxed halos are poorly captured by the NFW fitting formula (e.g. \citealt{Neto}). Because halo concentrations are clearly affected by transient departures from equilibrium, we only consider relaxed halos in the remainder of this work.
 
\subsection{Comparison with previous studies}

In this section we compare the $c-M$ relations of the most recent studies on dark matter halo concentrations, \citet{vandenBosch14} (hereafter, vdB14), \citet{Diemer14b} (hereafter, DK14), \citet{Dutton14} (hereafter, DM14) and \citet{Ludlow14} (hereafter, L14), with the model presented in this work. 

vdB14 used the $c-M$ relation of \citet{Zhao09} (obtained from fits of a full halo sample from numerical simulations) and adjusted the parameters by fitting it to the $c-M$ relation of the full halo sample from the Bolshoi simulations. vdB14 assumed the Bolshoi cosmology (consistent with WMAP7, \citealt{Komatsu11}), but they made use of a semi-analytic model to scale their model to any cosmology. We assume the Planck cosmology and use the publicly available code of vdB14 to calculate their $c-M$ relation. DK14 obtained a concentration model given by a best-fit seven parameter function of peak height ($\nu$) and slope of the linear power spectrum. They considered their full halo sample and extended their model to make it suitable for any cosmology. Finally, DM14 followed the evolution of the concentration of relaxed dark matter halos from a series of $N$-body simulations that assumed the Planck cosmology. DM14 fitted a power-law to the $c-M$ relation and restricted their analysis to relaxed halos only.

The left panel of Fig.~\ref{comparison} shows a comparison of our $c-M$ model (solid lines) to the model of vdB14. To compare with vdB14, we predicted the concentrations using the analytic expression for the MAH assuming the Planck cosmology (shown in Fig.~\ref{Mz12_LCDM}) and a $z_{-2}-c$ relation with a constant of proportionality of 850 instead of the value 887 used for the WMAP5 cosmology (see Appendix~\ref{cosmo_dependence} for a discussion of the cosmology dependence of our model). We find broad agreement with the relation of vdB14 only at $z=1$ and $2$. In their work, \citet{vandenBosch14} used the \citet{Zhao09} model which assumes that $c$ never drops below 4 at high redshift.

The middle-panel of Fig.~\ref{comparison} shows the DK14 $c-M$ relation calculated assuming the Planck cosmology. As they included their entire sample of halos for their $c-\nu$ relations, they obtained an upturn at the high-mass end at all redshifts. We find that our model predicts concentrations that are a factor of 1.2 larger just before the high-mass upturn. Finally, the right-panel of Fig.~\ref{comparison} shows reasonable agreement between our model and the DM14 $c-M$ relation for $z=0, 1, 2$ and $3$ although the results diverge at low masses. In their work, DM14 fitted a power-law, $c\propto M^{\alpha}$, to the $c-M$ relation at all redshifts, and found that the slope, $\alpha$, increases from $-0.1$ at $z=0$, to $0.03$ at $z=5$. As they restricted their halo sample to relaxed halos, they did not obtain a significant upturn at the high-mass end of the $c-M$ relation.

\begin{figure*} 
  \begin{center}
  \includegraphics[angle=0,width=0.8\textwidth]{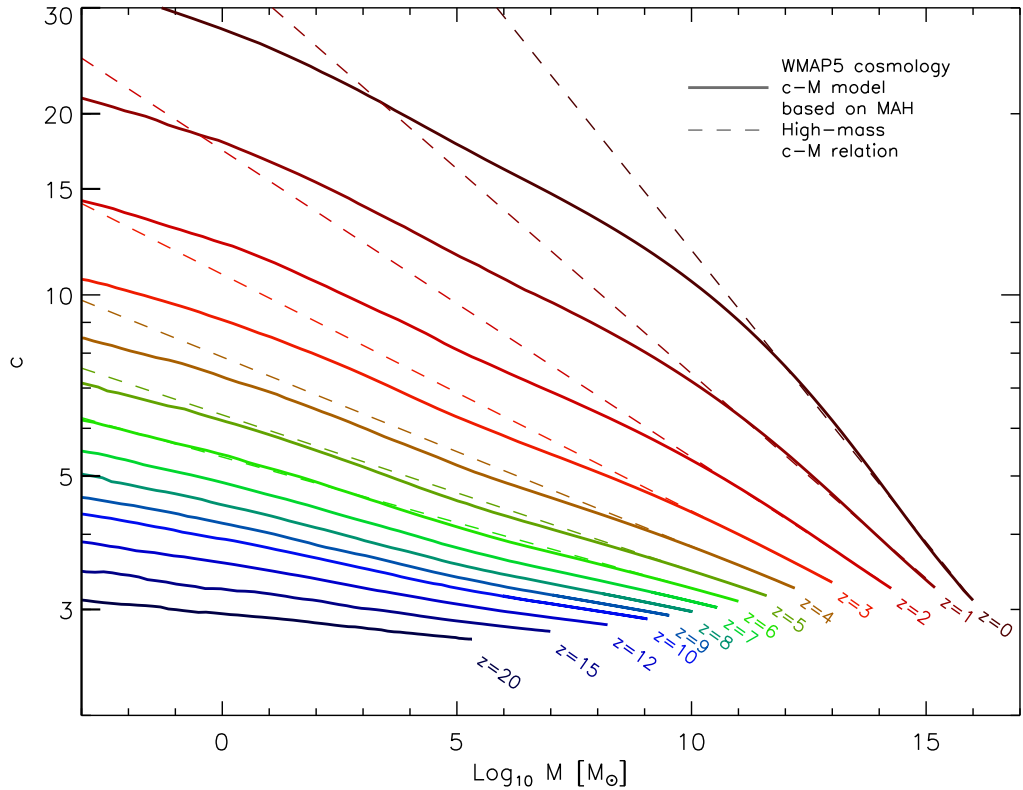}
  \end{center}
  \caption{Predicted concentration $-$ mass relation for the WMAP5 cosmology over a wide range of halo masses (${\log_{10}M/\Msol=[-2 ,16]}$) and redshifts (${z=0-20}$). The solid lines correspond to our $c-M$ model obtained from the halo MAH as described in Section~\ref{cM_model}. The lines are coloured as a function of redshift as indicated. The dashed lines correspond to power-law fits to the high-mass $c-M$ relation.}
\label{cvsM_big}
\end{figure*}

Fig.~\ref{comparison} shows that the physically motivated model presented in this work yields $c-M$ relations that are generally in agreement with previous results. However, the important improvement with respect to previous works is that we are presenting a physical analytic model that can then be extrapolated to very low-masses, and is suitable for any cosmology.

The model for dark matter halo concentrations presented in this work strongly relies on the relation $\langle\rho(r_{-2})\rangle-\rho_{\rm{crit}}(z_{-2})$, which supports the idea that halos grow inside-out. This relation was introduced in \cite{Ludlow13} and explored in L14, who recently presented a related model for the concentration-mass relation. In their work, L14 used the average MAHs from \cite{van} and \cite{Zhao09} that begin at $z_{i}=0$\footnote{L14 $c-M$ model used MAHs from \cite{van} and \cite{Zhao09} to show specific examples on how to construct a $c(M, z)$ relation for a given MAH, but any MAH model can be used.}. They fitted the halo MAHs, written as $M(\rho_{\rm{crit}})$, with the NFW profile expressed in terms of the enclosed density. They looked for a correlation between the concentration parameter $c_{\rm{MAH}}$, that results from an NFW fit to the halo MAH, and the concentration parameter of the halos density profile, $c_{\rm{NFW}}$, and used the best-fitting relation to predict $c_{\rm{NFW}}$ from $c_{\rm{MAH}}$. L14 and this work use the same formation redshift definition to connect concentrations with halo MAHs. L14 used the $\langle\rho(r_{-2})\rangle-\rho_{\rm{crit}}(z_{-2})$ relation to find the $c_{\rm{MAH}}-c_{\rm{NFW}}$ relation, whereas in this work we used the analytic MAH model to define formation redshifts and used the $\langle\rho(r_{-2})\rangle- \rho_{\rm{crit}}(z_{-2})$ relation to predict concentrations. Although there is good agreement between L14 and our $c-M$ relation at $z=0$, there are differences in the relations at high$-z$, e.g. a factor of 1.2 difference between the concentrations of a $10^{10}h^{-1}\rm{M}_{\odot}$ halo at $z=2$ ($c\sim 5.25$ versus $c_{\rm{L14}}\sim 6.3$), and a factor of 1.58 for a $10^{5}h^{-1}\rm{M}_{\odot}$ halo at $z=2$ ($c\sim 7.95$ versus $c_{\rm{L14}}\sim 12.58$), for the WMAP5 cosmology. Those differences are mainly due to the different MAH models. Since the $\langle\rho(r_{-2})\rangle-\rho_{\rm{crit}}(z_{-2})$ relation is essentially equivalent to the $c_{\rm{MAH}}-c_{\rm{NFW}}$ relation, we expect L14 and our semi-analytic model to give consistent results if the same MAH model is used. We believe however that we have improved upon the L14 $c-M$ model by combining the $\langle\rho(r_{-2},z_{i})\rangle-\rho_{\rm{crit}}(z_{-2},z_{i})$ relation with an analytical MAH model, $M(z,z_{i})$, that begins at any redshift $z_{i}$, and allows a detailed analysis of the redshift dependence of the $c(M,z)$ relation for relaxed halos. Another important difference is the tentative evidence for a cosmology dependence in the $\langle\rho(r_{-2},z_{i})\rangle-\rho_{\rm{crit}}(z_{-2},z_{i})$ relation (for a discussion see Appendix~\ref{cosmo_dependence}).

\subsection{Extrapolation to low halo masses and high redshifts}

Because our semi-analytic model for halo concentration is physical, rather than a purely empirical fit to the simulation results, we can use it to extrapolate beyond the mass and redshift ranges spanned by our simulations, assuming that the $z_{-2}-c$ relation given by eq. (\ref{zf_relation}) holds. Fig.~\ref{cvsM_big} shows the predicted concentration-mass relation for a wide range of halo masses (${\log_{10}M/\Msol=[-2 ,16]}$) and redshifts (${z=0-20}$). The dashed lines correspond to the high-mass power-law $c-M$ relations at low redshift. These are included to aid the comparison of the slopes of the $c-M$ relation in the high- and low-mass regimes. There is a clear `break' in the $z=0$ $c-M$ relation. For $M>10^{12}\Msol$ concentration scales as $c\propto M^{-0.083}$, whereas at $M<10^{9}\Msol$ it scales as $c\propto M^{-0.036}$. The change of slope around these halo masses is substantial up to $z=3-4$. However, at $z>4$ there is no `break' in the $c-M$ relation. In Section~\ref{c_evolution}, we provide a tentative explanation for the physical origin of the break in the $c-M$ relation.

We provide fitting functions for the $c-M$ relation in the high-$z$ and low-$z$ regimes. The following expression is suitable for the low-redshift regime ($z\le 4$) and at all halo masses,

\begin{eqnarray}\label{c_lowz}
\log_{10} c &=& \alpha+\beta\log_{10}(M/\Msol)[1+\gamma(\log_{10}M/\Msol)^{2}],\\\nonumber
\alpha &=& 1.62774-0.2458(1+z)+0.01716(1+z)^{2},\\\nonumber
\beta &=& 1.66079+0.00359(1+z)-1.6901(1+z)^{0.00417},\\\nonumber
\gamma &=& -0.02049+0.0253(1+z)^{-0.1044}.
\end{eqnarray}

In the high-redshift regime the $c-M$ relation can be fitted using only two parameters. The following expression is suitable for $z>4$ and at all halo masses,

\begin{eqnarray}\label{c_highz}
\log_{10} c &=& \alpha+\beta\log_{10}(M/\Msol),\\\nonumber
\alpha &=& 1.226-0.1009(1+z)+0.00378(1+z)^{2},\\\nonumber
\beta &=& 0.008634-0.08814(1+z)^{-0.58816}.
\end{eqnarray}

\begin{figure}  
  \begin{center}
  \includegraphics[angle=0,width=0.46\textwidth]{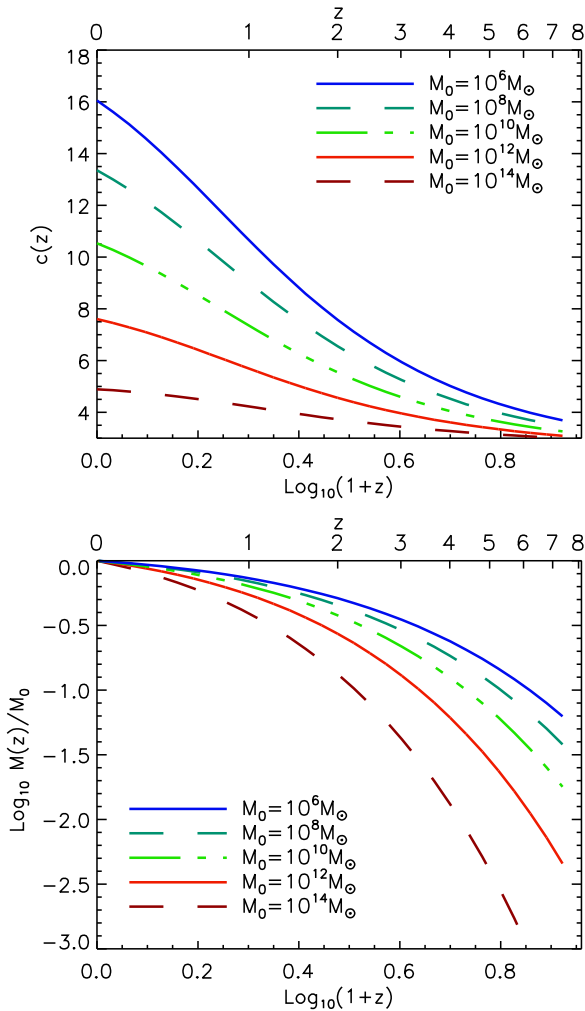}\\
  \end{center}
  \caption{{\it{Top panel}}: evolution of the concentrations of halos that at $z=0$ have masses of $M_{0}=10^{6}, 10^{8}, 10^{10}, 10^{12}$ and $10^{14}\Msol$, as indicated in the legends. {\it{Bottom panel}}: MAHs of halos of the same masses as in the top panel.}
  \label{cMH_figa}
\end{figure}

\begin{figure}  
  \begin{center}
  \includegraphics[angle=0,width=0.46\textwidth]{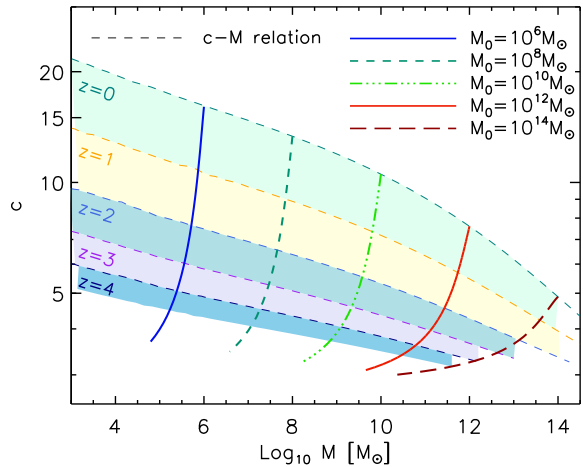}\\
  \end{center}
  \caption{Dashed curves show the concentration$-$mass relation at various redshifts $z$. The solid and dashed lines show the $c-M$ evolution of halos whose $c(z)$ and $M(z)$ are shown in the top and bottom panels of Fig.~\ref{cMH_figa}, respectively. All results were obtained using the physical model described in Sections \ref{Halo_history_modelling} and \ref{cM_model}.}
  \label{cMH_figb}
\end{figure}

The above fitting functions have been calculated assuming the WMAP5 cosmology. Appendix~\ref{cosmo_dependence} provides a series of best-fitting relations for the Planck cosmology, as well as a short discussion of the cosmology dependence of the $c-M$ relation presented in this work. In addition, Appendix~\ref{script_details} provides a description of a simple code (available  for download in IDL and Python) that computes concentrations, MAHs and accretion rates as a function of redshift for any cosmology.

The $c-M$ model presented in this work predicts a concentration of $c=3$ for microhalos of $10^{-7}\Msol$ at $z=31$, in agreement with simulations of microhalo formation from \citet{Anderhalden} and \citet{Ishiyama}. In their work, \citet{Anderhalden} compared the empirical $c-M$ relations from \citet{Bullock} and \citet{Maccio08} to their simulation outputs, and concluded that extrapolating simple power-law approximations to typical microhalo scales results in an  overestimation of $c$ by up to a factor of 10 at low$-z$. We also find large differences between extrapolations of the high-mass power-law fits to low-masses and the predictions of our physical model, as can be seen by comparing the dashed and solid lines in Fig.~\ref{cvsM_big}. This impacts calculations of the dark matter (DM) annihilation signal boost (see e.g. \citealt{Sanchez}) and the power from DM annihilation (see e.g. \citealt{Mack}), which make extensive use of $c-M$ relations at various redshifts. In Section~\ref{DM_annihilation_signal} we analyse this issue in more detail by calculating the power from DM annihilation, assuming either the \citet{Duffy08} $c-M$ relation (an empirical power-law fit for high halo masses) or the $c-M$ relation from this work.

\section{Evolution of the concentration}\label{c_evolution}

In this section we use our semi-analytic model to investigate the evolution of concentration and the effects that determine the slope of the $c-M$ relation. The top panel of Fig. \ref{cMH_figa} shows the evolution of the concentration of halos that have masses of $M_{0}=10^{6}, 10^{8}, 10^{10}, 10^{12}$ and $10^{14}\Msol$ at $z=0$. The bottom panel shows the corresponding halo MAHs normalized to the final halo mass at $z=0$ ($M_{0}$). We computed $c(z)$ and $M(z)$ following the models described in Sections~\ref{cM_model} and \ref{MAHmodel_highz}, respectively. 

In Paper I, we used EPS theory to show that the MAH of all halos can be described by the expression ${M(z)=M_{0}(1+z)^{\alpha}e^{\beta z}}$, where the exponential is due to the fast growth at high$-z$ and the power-law due to the slow growth at low$-z$. In addition, the parameters $\alpha$ and $\beta$ depend on halo mass. As a result, MAHs of halos larger than $10^{11}\Msol$ are mainly characterized by an exponential growth, whereas lower-mass halos exhibit a MAH closer to a power law, as can be seen in the bottom panel of Fig.~\ref{cMH_figa}.

Comparing the coloured curves in the top and bottom panels of Fig.~\ref{cMH_figa}, we see an interesting relation between the evolution of $c$ and the corresponding MAH. Dark matter halos with a small growth rate are {\it{appear}} to contract, and so their concentrations grow rapidly. This can be understood as follows. At low redshift, during the dark energy dominated epoch, $M(z)$ of low-mass halos is characterized by a power law (\citealt{PaperI}). During this epoch, there is a drop in the accretion and merger rates of small halos, and the halo mass increases due to the evolution of the reference density used in the spherical overdensity definition of the halo ($\rho_{\rm{crit}}(z)$ in this case). This so-called {\it{pseudo-evolution}} of the halo mass is thus driven by the halo mass definition rather than the accretion of new material (see \citealt{Diemer13} and references therein). The pseudo-evolution of the halo mass gives the impression that concentrations are increasing because of contraction of the bound cores (\citealt{Tasitsiomi,Zhao03,LuYu,Li08,vandenBosch14}), when in fact the core radius remain constant. Indeed, the evolution of the $c-M$ relation has been shown to be sensitive to the definition of halo mass (e.g. \citealt{Duffy08}). If we assume that $r_{-2}$ is constant in the redshift range $z=0-1$, then the increase in $r_{200}$ due to the drop in $\rho_{\rm{crit}}$ gives the approximate increase in the concentration values. We find $\frac{c(z=0)}{c(z=1)}=\frac{r_{200}(z=0)}{r_{200}(z=1)}=\left(\frac{\rho_{\rm{crit}}(z=1)}{\rho_{\rm{crit}}(z=0)}\right)^{1/3}\sim 1.4$, in agreement with the increase in concentration of a $M_{0}=10^{6}\Msol$ halo.

In the high-redshift regime ($z\gg 1$, matter-dominated epoch), the halo MAH is mainly characterized by exponential growth. During this time, concentrations grow by a factor of 2 (from $z=8$ to $z=2$) for a $M_{0}=10^{6}\Msol$ halo, decreasing to a factor of 1.08 (from $z=8$ to $z=2$) for a  $M_{0}=10^{14}\Msol$ halo. The pseudo-evolution of the halo mass is negligible in comparison with the high accretion rates, and the core radius increases simultaneously with the virial radius, hence the concentration hardly grows.

In the case where the halo mass history is characterized by exponential growth at all $z$, representing the situation of a universe with no dark energy but $\Omega_{\rm{m}}\le 1$, we find that concentrations do not reach such large values at $z=0$. We thus conclude that the evolution of the concentration is indirectly affected by the accelerated expansion of the Universe through the MAH and the halo mass definition.

Next, we analyse how the evolution of the concentration determines the change in slope of the $c-M$ relation. Fig.~\ref{cMH_figb} shows $c-M$ relations at various redshifts (dashed lines), and the $c-M$ evolution of halos with $M_{0}=10^{6}, 10^{8}, 10^{10}, 10^{12}$ and $10^{14}\Msol$ (solid lines). From this figure, we see that the `break' in the low-redshift $c-M$ relation that occurs at $M\sim 10^{11}\Msol$ is produced by the change in the halo MAH.

As mentioned, $M(z)$ changes from being dominated by exponential growth for high-mass halos, to power-law growth for low-mass halos. It is natural to ask why the break in $M(z)$ (and consequently in $c-M$) occurs at $\sim 10^{11}\Msol$. The answer is given by the rms of the linear theory density perturbation field, $\sigma$, that determines at which halo mass $M(z)$ changes from power-law dominated to exponential dominated. Since $\sigma\gg1$ at low masses ($\ll 10^{11}\Msol$) and $\sigma\ll 1$ at high masses ($\gg 10^{11}\Msol$), the low values of $\sigma$ at high masses increase the value of the $\beta$ parameter in the exponential function of the $M(z)$ model, and thus $M(z)$ is mostly dominated by the exponential growth. As a result, higher mass halos increase their mass faster, their inner cores increase with the virial radius, and their concentrations do not grow as rapidly. The different growth rate of the concentrations produced by the change in the halos MAH, creates the `break' in the $c-M$ relation. 

Therefore, the break can be understood as being produced by the varying power in the density perturbations through the halo MAH, where the MAH of low-mass halos at $z<1$ is mostly driven by pseudo-evolution. The break is less prominent at $z>1$, because at higher redshifts $M(z)$ is mostly exponential for all halo masses (see eq.~\ref{alpha_tilde}, $\tilde{\alpha}\rightarrow 0$ for $z_{i}>0$ due to the growth factor and Fig.~\ref{MAH_z0}).

\begin{figure*} 
\begin{center}
\includegraphics[angle=0,width=\textwidth]{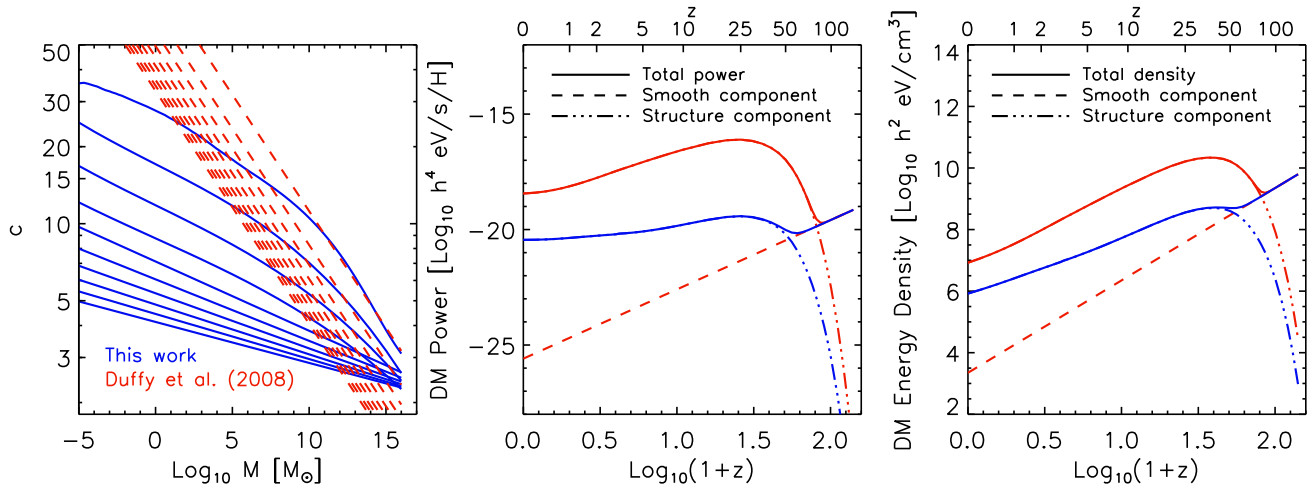}\\
\caption{{\it{Left panel:}} Concentration-mass relation at various $z$ ($z=0-10$, top to bottom) from this work (blue solid lines) and from Duffy et al. (2008) (red dashed lines). At $z=10$ and at a mass-scale of $1\Msol$ concentrations of Duffy et al. (2008) are a factor of 10 larger than concentrations predicted by this work, and a factor of 40 for a $10^{-9}\Msol$ halo. {\it{Middle panel:}} DM annihilation power per hydrogen nucleus as a function of redshift. The dashed line corresponds to the smooth component of the power and the dot-dashed lines to the structure component. The solid lines show the sum of the two components (structure+smooth). The red lines correspond to the power assuming the \citet{Duffy08} $c-M$ relation, whereas the blue lines correspond to the power assuming the $c-M$ relation from this work. {\it{Right panel:}} as the middle panel, but showing the effective DM energy density as a function of redshift.}
\label{DM}
\end{center}
\end{figure*}


\section{Implications for the dark matter annihilation signal}\label{DM_annihilation_signal}

Dark matter (DM) particles are predicted to self-annihilate into Standard Model particles, thus injecting energy into the surrounding medium (e.g. \citealt{Furnaletto2006}). In this section we calculate the DM annihilation rate per unit volume produced by a smooth density field of DM (dominant before structure formation) and by cosmic structures (halos and microhalos). We separate DM into a smooth and structure component because the spatial distribution of mass is almost completely smooth at very early cosmic times. Later gravitational instability causes overdensities to grow, until micro DM halos form. We follow \citet{Cirelli} and \citet{Mack} in this calculation, and obtain the DM energy density and mean power from DM annihilation, assuming the physical $c-M$ relation from this work and from extrapolations of the fits to simulations from \citet{Duffy08}. Below we briefly describe the calculation of the DM annihilation rate produced by cosmic structures.


The DM annihilation rate per unit volume results from the sum of two parts, a structure contribution and a smooth contribution. The smooth contribution, dominant before structure formation, $z\gtrsim 100$, can be written as

$$R^{\rm{smooth}}(z)= \frac{\langle\sigma\nu\rangle}{2m_{\chi}^{2}}\rho^{2}_{\rm{DM,0}}(1+z)^{6}.$$

\noindent Where $m_{\chi}$ is the mass of the DM particle,  $\langle\sigma\nu\rangle$ the self-annihilation cross-section, and $\rho_{\rm{DM,0}}$ the smooth DM density today, $\rho_{\rm{DM,0}}=\Omega_{\rm{DM}}\rho_{\rm{crit}}$.

The DM annihilation rate per unit volume due to halos, $R$, is given by

\begin{equation}\label{integral}
R(z)=\frac{\langle\sigma v \rangle}{2m_{\chi}^{2}}\int dM\frac{dn}{dM}(z,M)(1+z)^{3}\int dr 4\pi r^{2}\rho^{2}(r,M).
\end{equation}

\noindent Here $m_{\chi}$ is the mass of the dark matter particle and $\langle\sigma v \rangle$ is the average annihilation cross section, which we assume to be $100$ GeV and $10^{-26}\rm{cm}^{3}\rm{s}^{-1}$, respectively (e.g. \citealt{Aprile2012}). For the halo mass function, $\frac{dn}{dM}(z,M)$, we adopt the expression from \citet{Reed}. For $\rho(r,M)$ we use the NFW density profile. We use $M_{200}$ as the halo mass definition. 

Next, we calculate the effective DM density from structure formation, defined as 

\begin{eqnarray}
\rho_{\rm{DM}}^{\rm{eff}}(z) &\equiv&\rho_{\rm{DM},0}(1+z)^{3}R_{i}(z),\\
R_{i}(z) &\equiv & \left(\frac{2m_{\chi}^{2}}{\langle\sigma v \rangle}R(z)\right)^{1/2},
\end{eqnarray}

\noindent where $\rho_{\rm{DM},0}$ is the average DM density today, ${\rho_{\rm{DM},0}=\Omega_{\rm{DM}}\rho_{\rm{crit},0}}$, with $\Omega_{\rm{DM}}h^{2}=0.11$. In this calculation we assume the WMAP5 cosmology in order to facilitate a comparison with models using the extrapolation of the Duffy et al. (2008) power-law fit to the $c-M$ relation predicted by WMAP5 $N$-body simulations.

Similarly, we calculate the averaged volume power, per hydrogen nucleus, produced from DM annihilation events as

\begin{equation}
P(z)=2m_{\chi}c^{2}\frac{R(z)}{n_{\rm{H}}(z)},
\end{equation}

\noindent with $n_{\rm{H}}(z)= \Omega_{\rm{b}}\rho_{\rm{crit},0}(1-Y_{\rm{p}})(1+z)^{3}/m_{\rm{H}}$,\footnote{$\Omega_{\rm{b}}$ is the present day baryon density parameter, $Y_{\rm{p}}=0.24$ the primordial mass fraction of helium and $m_{\rm{H}}$ the proton mass.} the number density of hydrogen.

\subsection{Implications}

Several of the models that have been used to predict the DM annihilation signal (see e.g. \citealt{Lavalle,Pieri,Pinzke}) have extrapolated $c-M$ relations, obtained from power-law fits to simulation results, to mass far below the resolution limit of the simulations. These power-law extrapolations assign huge concentrations to the smallest halos, thus increasing the DM annihilation power. In this section we explore how our physically motivated $c-M$ relation, which flattens towards low-masses at low-redshift, affects the DM annihilation power when comparing it to the power calculated using an extrapolation of the \citet{Duffy08} power-law $c-M$ relation (the results are similar for other published power-law fits to the $c-M$ relation from simulations).

The left panel of Fig.~\ref{DM} shows a comparison between the concentration-mass relations at various redshifts ($z=0-10$) from this work (blue solid lines) and of \citet{Duffy08} (red dashed lines). For $z=0$ there is good agreement at the high-mass end between both relations. However, at $z=10$ and at a mass-scale of $1\Msol$, the concentrations of \citet{Duffy08} are a factor of 10 larger than the concentrations predicted by this work. For mass-scales of $10^{-9}\Msol$ the difference is a factor of 40. In the middle and right panels of Fig.~\ref{DM} we investigate the effects of these different $c-M$ relations, showing the DM annihilation power per hydrogen nucleus and the effective DM energy density, respectively, as a function of redshift. In the middle panel, the dashed line corresponds to the smooth DM component of the power whereas the dot-dashed lines correspond to the structure component. The solid line shows the sum of the two components (structure+smooth). In each case the red lines correspond to the power assuming the \citet{Duffy08} $c-M$ relation, whereas the blue lines correspond to the power assuming the $c-M$ relation from this work. The change in $c-M$ affects the normalization of the power as well as the redshift at which structures begin to dominate. Lower concentrations result in lower central densities. Since the annihilation rate per unit volume, $R$, scales as $\rho^{2}$, it is clear that $R$ should decrease accordingly. At $z=0$, the DM annihilation power that assumes the $c-M$ relation predicted by this work is two orders of magnitude lower than the power obtained by extrapolating the \citet{Duffy08} $c-M$ relation. In addition, the higher concentrations predicted by the \citet{Duffy08} $c-M$ relation imply that halos dominate the power over the smooth DM density component at higher redshifts. Adopting the $c-M$ relation from this work results in the power from structures starting to dominate at $z\approx 50$ rather than at $z\approx 85$ (in agreement with \citealt{Mack} and \citealt{Ng2014}). This lower redshift of structure formation dominating over the smooth component could have important implications for searches of the `Dark Ages' by radio telescopes (e.g. \citealt{Pritchard}).
 
\section{Discussion}\label{Discussion}

\begin{figure} 
 \centering
 \includegraphics[angle=0,width=0.48\textwidth]{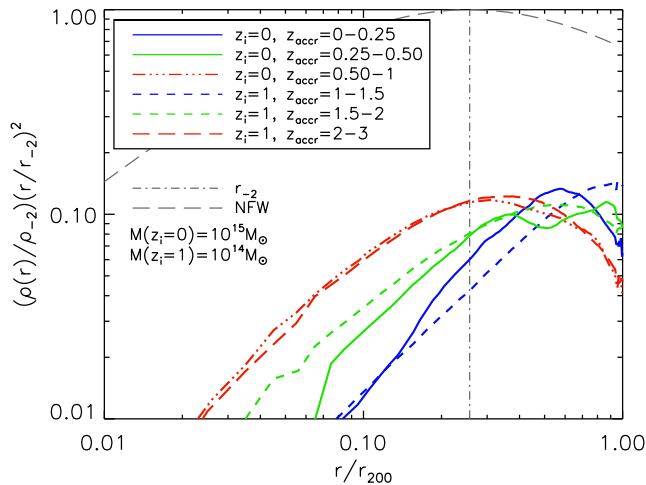}
 \caption{Distribution of particles within $r_{200}$. We show the particle distribution of two different halo samples, the first sample contains halos of $10^{15}\Msol$ identified at $z_{i}=0$ (solid lines) and the second sample contains halos of $10^{14}\Msol$ identified at $z_{i}=1$ (dashed lines). The solid grey line corresponds to a NFW density profile with concentration $c=4$ and the vertical dashed line to the corresponding $r_{-2}$ radius. The different colours indicate the redshift ($z_{\rm{accr}}$) during which the particles were accreted on to the halos (note that the same colour corresponds to different redshift ranges for different $z_{i}$).}
 \label{discussion_plot}
\end{figure}

Dark matter halo concentrations have recently been the subject of extensive analysis due to the controversial reports of an upturn at the high-mass end of the relation (\citealt{Munoz,Prada,Klypin14,Diemer14b}). However, the semi-analytic model for dark matter halo concentrations presented in this work does not predict such an upturn. In this section, we review the main assumptions that the model relies on and discuss the plausibility of the existence of the upturn.

First, the model assumes that the halo density profile is described by the NFW profile at all times. Although it is known that the Einasto profile (\citealt{Einasto}) is more accurate than the NFW profile (\citealt{Gao08}), it has an extra `shape' parameter that complicates the fitting procedure and affects the concentration. Also, the residuals from the systematic deviations from the NFW shape are generally smaller than $10\%$ and the NFW concentrations only differ by $10-20\%$ from Einasto fit and the velocity profile fit (as recently discussed by \citealt{Dutton14} and \citealt{Klypin14}). We thus conclude that using the NFW profile to predict densities is not a major determinant in the model. 

Secondly, the model depends on the calibration of the $\rho_{\rm{crit}}(z_{-2})-\langle\rho\rangle(<r_{-2})$ relation, which implies that halo formation is an `inside out' process, where the central part of a dark matter halo (contained within $r_{-2}$) forms first, and later accretion and mergers increase the mass and size of the halo without adding much material to its inner regions (\citealt{Huss,Wang09}). We test the assumption of `halo formation' of our model. We analyse the distribution of particles within $r_{200}$, and differentiate the particles according to the period of time during which they were accreted. Fig.~\ref{discussion_plot} shows the radial distribution of particles of two different halo samples. The first sample contains halos of $10^{15}\Msol$ identified at $z_{i}=0$ (which are formed at $z_{-2}\approx 1$) and the second sample contains halos of $10^{14}\Msol$ identified at $z_{i}=1$ (formed at $z_{-2}\approx 2$). We analysed halos of different masses but focused on massive halos, because their large radial velocities makes them more likely to contain recently accreted particles in their inner regions, and they are located in the upturn in the $c-M$ relation.

Fig.~\ref{discussion_plot} shows the NFW density profile (in grey solid line) and the $r_{-2}$ radius (in vertical dashed line) for a concentration of $c=4$. The figure also shows the distribution of particles at $z_{i}=0$ (solid coloured lines) and at $z_{i}=1$ (dashed coloured lines). The different colours indicate the redshift ($z_{\rm{accr}}$) during which the particles were accreted on to the halos. The blue lines show that recently accreted particles are distributed around $r_{200}$ and that only a tiny fraction ($<2\%$ of the total, in the two cases), reside in the inner parts of the halo. In the case of the distribution of particles at $z_{i}=0$, $2.5\%$ of the total particles are in regions within $r_{-2}$ after being accreted during $z_{\rm{accr}}=0.25-0.5$, and $4.5\%$ during $z_{\rm{accr}}=0.5-1$. The same behaviour is observed in halos of different masses identified at higher redshifts. We find that $8\%$ of all particles accreted after the halo has formed are in the centre, not enough to significantly alter the mass within $r_{-2}$ so as to increase concentration. We then find the halo formation assumption that the model relies on to be valid.

Finally, in the calibration of the $\rho_{\rm{crit}}(z_{-2})-\langle\rho\rangle(<r_{-2})$ relation, we only consider relaxed halos. The selection conditions generally used to differentiate relaxed halos from unrelaxed (\citealt{Maccio07,Neto}) have recently been revisited by \citet{Klypin14}. These conditions include the virial parameters ($2K/|W|-1$, where $K$ and $W$ are the kinetic and potential energies), the offset parameter $X_{\rm{off}}$ (distance between the potential minimum and the center of mass), and the spin parameter. In their work, \citet{Klypin14} argued that the virial equilibrium condition is too simplistic and needs to include the effects of the surface pressure and external forces. They applied these corrections to the virial parameters and selected halos that had previously been rejected. As a result, they obtained an upturn in the high-mass end of the $c-M$ relation and claimed that the large concentration of massive halos is due to their infall velocities, which are more radial and result in deeper penetration of infalling mass into the halo that reaches the inner parts. In this work we selected relaxed halos using only the condition that $X_{\rm{off}}<0.07$ following \citet{Duffy08} and \citet{Neto}, who found that this simple criterion resulted in the removal of the vast majority of unrelaxed haloes. We did not use any additional criteria and did not find any upturn at high halo masses, but concluded that the strong flattening of the $c-M$ relation at high redshift is due to unrelaxed halos (Fig.~\ref{relaxed_comparison}). We cannot say we disagree with \citet{Klypin14} regarding the relaxation conditions, because our simulations do not have sufficiently large box sizes to model a large sample of the massive highly unrelaxed halos that likely `shape' the upturn.
 
\section{Conclusion}\label{Conclusion}

In this paper, we have linked the concentration of a halo to its MAH. We extended the analytic framework presented in Paper I to show that the halo mass history $\tilde{M}(z,M(z_{i}),z_{i})$ of a halo with mass $M(z_{i})$ at $z_{i}$ can be described by

$$\tilde{M}(z,M(z_{i}),z_{i})= M(z_{i})(1+z-z_{i})^{\tilde{\alpha}}e^{\tilde{\beta}(z-z_{i})},$$

\noindent where $\tilde{\alpha}$ and $\tilde{\beta}$ are parameters that depend on $M(z_{i})$ and cosmology. We have compared the above formula to simulation outputs and the most recent empirical mass history model from the literature (\citealt{vandenBosch14}) and found generally reasonable agreement.

Building on the work by \citet{Ludlow14}, we presented a semi-analytic model for halo concentration as a function of halo mass and redshift. The model connects the analytic model for the MAH from Paper I to halo concentration through an empirical relation between concentration and formation redshift, obtained through fits to simulation data in Paper II. The formation redshift definition of \citet{Ludlow13} and \citealt{Ludlow14}, $z_{-2}$, defined as the redshift at which the mass of the main progenitor equals the mass enclosed within the scale radius of the NFW density profile at $z=z_{i}$, results in an inner halo structure that reflects the background density of the Universe at the time when the halo formed. 

The resulting $c-M$ relations were tested using $N$-body simulations, and compared to the most recent empirical $c-M$ relations from the literature (\citealt{vandenBosch14,Dutton14,Diemer14b}). The `upturn' at high masses seen by some studies (\citealt{Klypin2011,Prada,Dutton14,Diemer14b}) is not reproduced by our physically derived model which, however, only applies to relaxed halos. We analysed the $c-M$ relations obtained from the numerical simulations, where we differentiated between relaxed and unrelaxed halos. We found that the upturn is due to the inclusion of unrelaxed halos, supporting the previous claim of \citet{Ludlow12}.

We applied our model to a large range of halo mass (${\log_{10}M/\Msol=[-2 ,16]}$) and redshift (${z=0-20}$), and provided fits to the $c-M$ relations as well as numerical routines\footnote{Available at \href{https://bitbucket.org/astroduff/commah}{\it{https://bitbucket.org/astroduff/commah}} and \href{http://astro.physics.unimelb.edu.au/Research/Public-Data-Releases/COMMAH}{\it{http://astro.physics.unimelb.edu.au/}} in Research/Public-Data-Releases/COMMAH. See Appendix \ref{script_details} for a short description.} to compute concentrations and MAHs as a function of halo mass, redshift and cosmology. We caution the reader that baryonic processes will almost certainly increase the scatter in the $c-M$ relation and will modify the inner density profile (e.g. \citealt{Duffy10,Governato12,Teyssier13,Schaller}). 

Our model predicts a change in the slope of the $c-M$ relation at $z=0-3$ and a `break' in the $z\sim 0$ $c-M$ relation at a mass of $\sim 10^{11}\Msol$. We analysed the evolution of concentration and found that it increases more rapidly during the dark energy era, when the accretion rates of dark matter halos decrease due to the accelerated expansion of the Universe. We found that the break at a halo mass $\sim 10^{11}\Msol$ results from the change of the functional form of $M(z)$, that goes from being dominated by a power-law (for low-mass halos) to an exponential (for high-mass halos). This change in $M(z)$ is driven by the rms of the linear density perturbation field. Halos with mass $M\gg 10^{11}\Msol$, have $\sigma\ll 1$, are characterized by an exponential growth, and have lower concentrations as a result. Whereas halos with mass $M\ll 10^{11}\Msol$, have large $\sigma$, are characterized by a power-law growth. In this last case, there is a pseudo-evolution in the halo masses (i.e. mass growth due to the definition of the halo in terms of an overdensity criterion, \citealt{Diemer13}) and the core radius remains approximately constant, causing the concentrations to grow. The different growth rate of the concentrations at low and high mass produces the break in the $z\sim 0$ $c-M$ relation. This break is not so evident for $z>1$, because at higher redshifts $M(z)$ is mostly exponential for all halo masses ($\tilde{\alpha}\rightarrow 0$ for $z_{i}>0$ due to the growth factor), causing all concentrations to grow at approximately the same rate (as seen in Fig.~\ref{cMH_figa}).    

Finally, we addressed the impact of the $c-M$ relation presented in this work on predictions for the dark matter annihilation signal. We calculated the DM annihilation rate from cosmic structures and compared the results obtained by extrapolating the \citet{Duffy08} power-law fit (which is similar to other published fits to the results of simulations) to the rate obtained by using the $c-M$ relation predicted by our model. We found that the power from DM annihilation at $z=0$ is two orders of magnitude lower than the power obtained by extrapolating the Duffy et al. (2008) $c-M$ relation (in agreement with \citealt{Mack,Ng2014}).
 
\section*{Acknowledgements}

We are grateful to the OWLS team for their help with the simulations. We would also like to thank the referee, Aaron Ludlow, for many insightful comments and suggestions that greatly improved the manuscript. JSBW is supported by an Australian Research Council Laureate Fellowship. JS acknowledges support by the European Research Council under the European Union's Seventh Framework Programme (FP7/2007-2013)/ERC Grant agreement 278594-GasAroundGalaxies.  

\bibliography{biblio}

\begin{thebibliography}{}

\bibitem[\protect\citeauthoryear{{Anderhalden} \& {Diemand}}{{Anderhalden} \&
  {Diemand}}{2013}]{Anderhalden}
{Anderhalden} D.,  {Diemand} J.,  2013, \jcap, 4, 9

\bibitem[\protect\citeauthoryear{{Aprile} \& et al.}{{Aprile} \&
  et~al.}{2012}]{Aprile2012}
{Aprile} E.,  et al. 2012, Physical Review Letters, 109, 181301

\bibitem[\protect\citeauthoryear{{Bryan} \& {Norman}}{{Bryan} \&
  {Norman}}{1998}]{BryanNorman}
{Bryan} G.~L.,  {Norman} M.~L.,  1998, \apj, 495, 80

\bibitem[\protect\citeauthoryear{{Bryan}, {Kay}, {Duffy}, {Schaye}, {Dalla
  Vecchia} \& {Booth}}{{Bryan} et~al.}{2013}]{Bryan13}
{Bryan} S.~E.,  {Kay} S.~T.,  {Duffy} A.~R.,  {Schaye} J.,  {Dalla Vecchia} C.,
     {Booth} C.~M.,  2013, \mnras, 429, 3316

\bibitem[\protect\citeauthoryear{{Bullock}, {Kolatt}, {Sigad}, {Somerville},
  {Kravtsov}, {Klypin}, {Primack} \& {Dekel}}{{Bullock} et~al.}{2001}]{Bullock}
{Bullock} J.~S.,  {Kolatt} T.~S.,  {Sigad} Y.,  {Somerville} R.~S.,  {Kravtsov}
  A.~V.,  {Klypin} A.~A.,  {Primack} J.~R.,    {Dekel} A.,  2001, \mnras, 321,
  559

\bibitem[\protect\citeauthoryear{{Cirelli}, {Iocco} \& {Panci}}{{Cirelli}
  et~al.}{2009}]{Cirelli}
{Cirelli} M.,  {Iocco} F.,    {Panci} P.,  2009, \jcap, 10, 9

\bibitem[\protect\citeauthoryear{{Correa}, {Wyithe}, {Schaye} \&
  {Duffy}}{{Correa} et~al.}{2015a}]{PaperI}
{Correa} C.~A.,  {Wyithe} J.~S.~B.,  {Schaye} J.,    {Duffy} A.~R.,  2015a,
  \mnras, 450, 1514 (Paper I)

\bibitem[\protect\citeauthoryear{{Correa}, {Wyithe}, {Schaye} \&
  {Duffy}}{{Correa} et~al.}{2015b}]{PaperII}
{Correa} C.~A.,  {Wyithe} J.~S.~B.,  {Schaye} J.,    {Duffy} A.~R.,  2015b,
  \mnras, 450, 1521 (Paper II)

\bibitem[\protect\citeauthoryear{{Dalal}, {Lithwick} \& {Kuhlen}}{{Dalal}
  et~al.}{2010}]{Dalal}
{Dalal} N.,  {Lithwick} Y.,    {Kuhlen} M.,  2010, ArXiv e-prints:1010.2539

\bibitem[\protect\citeauthoryear{{Davis}, {Efstathiou}, {Frenk} \&
  {White}}{{Davis} et~al.}{1985}]{Davis}
{Davis} M.,  {Efstathiou} G.,  {Frenk} C.~S.,    {White} S.~D.~M.,  1985, \apj,
  292, 371

\bibitem[\protect\citeauthoryear{{Diemer} \& {Kravtsov}}{{Diemer} \&
  {Kravtsov}}{2015}]{Diemer14b}
{Diemer} B.,  {Kravtsov} A.~V.,  2015, \apj, 799, 108 (DK14)

\bibitem[\protect\citeauthoryear{{Diemer}, {More} \& {Kravtsov}}{{Diemer}
  et~al.}{2013}]{Diemer13}
{Diemer} B.,  {More} S.,    {Kravtsov} A.~V.,  2013, \apj, 766, 25

\bibitem[\protect\citeauthoryear{{Duffy}, {Schaye}, {Kay} \& {Dalla
  Vecchia}}{{Duffy} et~al.}{2008}]{Duffy08}
{Duffy} A.~R.,  {Schaye} J.,  {Kay} S.~T.,    {Dalla Vecchia} C.,  2008,
  \mnras, 390, L64

\bibitem[\protect\citeauthoryear{{Duffy}, {Schaye}, {Kay}, {Dalla Vecchia},
  {Battye} \& {Booth}}{{Duffy} et~al.}{2010}]{Duffy10}
{Duffy} A.~R.,  {Schaye} J.,  {Kay} S.~T.,  {Dalla Vecchia} C.,  {Battye}
  R.~A.,    {Booth} C.~M.,  2010, \mnras, 405, 2161

\bibitem[\protect\citeauthoryear{{Dutton} \& {Macci{\`o}}}{{Dutton} \&
  {Macci{\`o}}}{2014}]{Dutton14}
{Dutton} A.~A.,  {Macci{\`o}} A.~V.,  2014, \mnras, 441, 3359 (DM14)

\bibitem[\protect\citeauthoryear{{Einasto}}{{Einasto}}{1965}]{Einasto}
{Einasto} J.,  1965, Trudy Astrofizicheskogo Instituta Alma-Ata, 5, 87

\bibitem[\protect\citeauthoryear{{Eisenstein} \& {Hu}}{{Eisenstein} \&
  {Hu}}{1998}]{Eisenstein}
{Eisenstein} D.~J.,  {Hu} W.,  1998, \apj, 496, 605

\bibitem[\protect\citeauthoryear{{Furlanetto}, {Oh} \&
  {Pierpaoli}}{{Furlanetto} et~al.}{2006}]{Furnaletto2006}
{Furlanetto} S.~R.,  {Oh} S.~P.,    {Pierpaoli} E.,  2006, \prd, 74, 103502

\bibitem[\protect\citeauthoryear{{Gao}, {Navarro}, {Cole}, {Frenk}, {White},
  {Springel}, {Jenkins} \& {Neto}}{{Gao} et~al.}{2008}]{Gao08}
{Gao} L.,  {Navarro} J.~F.,  {Cole} S.,  {Frenk} C.~S.,  {White} S.~D.~M.,
  {Springel} V.,  {Jenkins} A.,    {Neto} A.~F.,  2008, \mnras, 387, 536

\bibitem[\protect\citeauthoryear{{Gao}, {Springel} \& {White}}{{Gao}
  et~al.}{2005}]{Gao}
{Gao} L.,  {Springel} V.,    {White} S.~D.~M.,  2005, \mnras, 363, L66

\bibitem[\protect\citeauthoryear{{Governato}, {Zolotov}, {Pontzen},
  {Christensen}, {Oh}, {Brooks}, {Quinn}, {Shen} \& {Wadsley}}{{Governato}
  et~al.}{2012}]{Governato12}
{Governato} F.,  {Zolotov} A.,  {Pontzen} A.,  {Christensen} C.,  {Oh} S.~H.,
  {Brooks} A.~M.,  {Quinn} T.,  {Shen} S.,    {Wadsley} J.,  2012, \mnras, 422,
  1231

\bibitem[\protect\citeauthoryear{{Huss}, {Jain} \& {Steinmetz}}{{Huss}
  et~al.}{1999}]{Huss}
{Huss} A.,  {Jain} B.,    {Steinmetz} M.,  1999, \apj, 517, 64

\bibitem[\protect\citeauthoryear{{Ishiyama}}{{Ishiyama}}{2014}]{Ishiyama}
{Ishiyama} T.,  2014, \apj, 788, 27

\bibitem[\protect\citeauthoryear{{Klypin}, {Yepes}, {Gottlober}, {Prada} \&
  {Hess}}{{Klypin} et~al.}{2014}]{Klypin14}
{Klypin} A.,  {Yepes} G.,  {Gottlober} S.,  {Prada} F.,    {Hess} S.,  2014,
  ArXiv e-prints:1411.4001

\bibitem[\protect\citeauthoryear{{Klypin}, {Trujillo-Gomez} \&
  {Primack}}{{Klypin} et~al.}{2011}]{Klypin2011}
{Klypin} A.~A.,  {Trujillo-Gomez} S.,    {Primack} J.,  2011, \apj, 740, 102

\bibitem[\protect\citeauthoryear{{Komatsu},  \& et al.}{{Komatsu}
  et~al.}{2009}]{Komatsu}
{Komatsu} E.,     et al. 2009, \apjs, 180, 330

\bibitem[\protect\citeauthoryear{{Komatsu} \& et al.}{{Komatsu} \&
  et~al.}{2011}]{Komatsu11}
{Komatsu} E.,  et al. 2011, \apjs, 192, 18

\bibitem[\protect\citeauthoryear{{Lavalle}, {Yuan}, {Maurin} \& {Bi}}{{Lavalle}
  et~al.}{2008}]{Lavalle}
{Lavalle} J.,  {Yuan} Q.,  {Maurin} D.,    {Bi} X.-J.,  2008, \aap, 479, 427

\bibitem[\protect\citeauthoryear{{Li}, {Mo} \& {Gao}}{{Li} et~al.}{2008}]{Li08}
{Li} Y.,  {Mo} H.~J.,    {Gao} L.,  2008, \mnras, 389, 1419

\bibitem[\protect\citeauthoryear{{Lu}, {Mo}, {Katz} \& {Weinberg}}{{Lu}
  et~al.}{2006}]{LuYu}
{Lu} Y.,  {Mo} H.~J.,  {Katz} N.,    {Weinberg} M.~D.,  2006, \mnras, 368, 1931

\bibitem[\protect\citeauthoryear{{Ludlow}, {Navarro}, {Angulo},
  {Boylan-Kolchin}, {Springel}, {Frenk} \& {White}}{{Ludlow}
  et~al.}{2014}]{Ludlow14}
{Ludlow} A.~D.,  {Navarro} J.~F.,  {Angulo} R.~E.,  {Boylan-Kolchin} M.,
  {Springel} V.,  {Frenk} C.,    {White} S.~D.~M.,  2014, \mnras, 441, 378 (L14)

\bibitem[\protect\citeauthoryear{{Ludlow}, {Navarro}, {Boylan-Kolchin}, {Bett},
  {Angulo}, {Li}, {White}, {Frenk} \& {Springel}}{{Ludlow}
  et~al.}{2013}]{Ludlow13}
{Ludlow} A.~D.,  {Navarro} J.~F.,  {Boylan-Kolchin} M.,  {Bett} P.~E.,
  {Angulo} R.~E.,  {Li} M.,  {White} S.~D.~M.,  {Frenk} C.,    {Springel} V.,
  2013, \mnras, 432, 1103

\bibitem[\protect\citeauthoryear{{Ludlow}, {Navarro}, {Li}, {Angulo},
  {Boylan-Kolchin} \& {Bett}}{{Ludlow} et~al.}{2012}]{Ludlow12}
{Ludlow} A.~D.,  {Navarro} J.~F.,  {Li} M.,  {Angulo} R.~E.,  {Boylan-Kolchin}
  M.,    {Bett} P.~E.,  2012, \mnras, 427, 1322

\bibitem[\protect\citeauthoryear{{Macci{\`o}}, {Dutton} \& {van den
  Bosch}}{{Macci{\`o}} et~al.}{2008}]{Maccio08}
{Macci{\`o}} A.~V.,  {Dutton} A.~A.,    {van den Bosch} F.~C.,  2008, \mnras,
  391, 1940

\bibitem[\protect\citeauthoryear{{Macci{\`o}}, {Dutton}, {van den Bosch},
  {Moore}, {Potter} \& {Stadel}}{{Macci{\`o}} et~al.}{2007}]{Maccio07}
{Macci{\`o}} A.~V.,  {Dutton} A.~A.,  {van den Bosch} F.~C.,  {Moore} B.,
  {Potter} D.,    {Stadel} J.,  2007, \mnras, 378, 55

\bibitem[\protect\citeauthoryear{{Mack}}{{Mack}}{2014}]{Mack}
{Mack} K.~J.,  2014, \mnras, 439, 2728

\bibitem[\protect\citeauthoryear{{Manrique}, {Raig}, {Salvador-Sol{\'e}},
  {Sanchis} \& {Solanes}}{{Manrique} et~al.}{2003}]{Manrique}
{Manrique} A.,  {Raig} A.,  {Salvador-Sol{\'e}} E.,  {Sanchis} T.,    {Solanes}
  J.~M.,  2003, \apj, 593, 26

\bibitem[\protect\citeauthoryear{{Mu{\~n}oz-Cuartas}, {Macci{\`o}},
  {Gottl{\"o}ber} \& {Dutton}}{{Mu{\~n}oz-Cuartas} et~al.}{2011}]{Munoz}
{Mu{\~n}oz-Cuartas} J.~C.,  {Macci{\`o}} A.~V.,  {Gottl{\"o}ber} S.,
  {Dutton} A.~A.,  2011, \mnras, 411, 584

\bibitem[\protect\citeauthoryear{{Navarro}, {Frenk} \& {White}}{{Navarro}
  et~al.}{1997}]{NFW97}
{Navarro} J.~F.,  {Frenk} C.~S.,    {White} S.~D.~M.,  1997, \apj, 490, 493 (NFW)

\bibitem[\protect\citeauthoryear{{Neistein}, {van den Bosch} \&
  {Dekel}}{{Neistein} et~al.}{2006}]{Neistein}
{Neistein} E.,  {van den Bosch} F.~C.,    {Dekel} A.,  2006, \mnras, 372, 933

\bibitem[\protect\citeauthoryear{{Neto}, {Gao}, {Bett}, {Cole}, {Navarro},
  {Frenk}, {White}, {Springel} \& {Jenkins}}{{Neto} et~al.}{2007}]{Neto}
{Neto} A.~F.,  {Gao} L.,  {Bett} P.,  {Cole} S.,  {Navarro} J.~F.,  {Frenk}
  C.~S.,  {White} S.~D.~M.,  {Springel} V.,    {Jenkins} A.,  2007, \mnras,
  381, 1450

\bibitem[\protect\citeauthoryear{{Ng}, {Laha}, {Campbell}, {Horiuchi},
  {Dasgupta}, {Murase} \& {Beacom}}{{Ng} et~al.}{2014}]{Ng2014}
{Ng} K.~C.~Y.,  {Laha} R.,  {Campbell} S.,  {Horiuchi} S.,  {Dasgupta} B.,
  {Murase} K.,    {Beacom} J.~F.,  2014, \prd, 89, 083001

\bibitem[\protect\citeauthoryear{{Pieri}, {Bertone} \& {Branchini}}{{Pieri}
  et~al.}{2008}]{Pieri}
{Pieri} L.,  {Bertone} G.,    {Branchini} E.,  2008, \mnras, 384, 1627

\bibitem[\protect\citeauthoryear{{Pinzke}, {Pfrommer} \&
  {Bergstr{\"o}m}}{{Pinzke} et~al.}{2011}]{Pinzke}
{Pinzke} A.,  {Pfrommer} C.,    {Bergstr{\"o}m} L.,  2011, \prd, 84, 123509

\bibitem[\protect\citeauthoryear{{Planck Collaboration} \& et al.}{{Planck
  Collaboration} \& et~al.}{2014}]{Planck}
{Planck Collaboration} et al. 2014, \aap, 571, A16

\bibitem[\protect\citeauthoryear{{Power}, {Navarro}, {Jenkins}, {Frenk},
  {White}, {Springel}, {Stadel} \& {Quinn}}{{Power} et~al.}{2003}]{Power03}
{Power} C.,  {Navarro} J.~F.,  {Jenkins} A.,  {Frenk} C.~S.,  {White} S.~D.~M.,
   {Springel} V.,  {Stadel} J.,    {Quinn} T.,  2003, \mnras, 338, 14

\bibitem[\protect\citeauthoryear{{Prada}, {Klypin}, {Cuesta}, {Betancort-Rijo}
  \& {Primack}}{{Prada} et~al.}{2012}]{Prada}
{Prada} F.,  {Klypin} A.~A.,  {Cuesta} A.~J.,  {Betancort-Rijo} J.~E.,
  {Primack} J.,  2012, \mnras, 423, 3018

\bibitem[\protect\citeauthoryear{{Pritchard} \& {Loeb}}{{Pritchard} \&
  {Loeb}}{2012}]{Pritchard}
{Pritchard} J.~R.,  {Loeb} A.,  2012, Reports on Progress in Physics, 75,
  086901

\bibitem[\protect\citeauthoryear{{Reed}, {Bower}, {Frenk}, {Jenkins} \&
  {Theuns}}{{Reed} et~al.}{2007}]{Reed}
{Reed} D.~S.,  {Bower} R.,  {Frenk} C.~S.,  {Jenkins} A.,    {Theuns} T.,
  2007, \mnras, 374, 2

\bibitem[\protect\citeauthoryear{{Salvador-Sol{\'e}}, {Vi{\~n}as}, {Manrique}
  \& {Serra}}{{Salvador-Sol{\'e}} et~al.}{2012}]{Salvador}
{Salvador-Sol{\'e}} E.,  {Vi{\~n}as} J.,  {Manrique} A.,    {Serra} S.,  2012,
  \mnras, 423, 2190

\bibitem[\protect\citeauthoryear{{S{\'a}nchez-Conde} \&
  {Prada}}{{S{\'a}nchez-Conde} \& {Prada}}{2014}]{Sanchez}
{S{\'a}nchez-Conde} M.~A.,  {Prada} F.,  2014, \mnras, 442, 2271

\bibitem[\protect\citeauthoryear{{Schaller}, {Frenk}, {Bower}, {Theuns},
  {Jenkins}, {Schaye}, {Crain}, {Furlong}, {Dalla Vecchia} \&
  {McCarthy}}{{Schaller} et~al.}{2015}]{Schaller}
{Schaller} M.,  {Frenk} C.~S.,  {Bower} R.~G.,  {Theuns} T.,  {Jenkins} A.,
  {Schaye} J.,  {Crain} R.~A.,  {Furlong} M.,  {Dalla Vecchia} C.,
  {McCarthy} I.~G.,  2015, \mnras, 451, 1247

\bibitem[\protect\citeauthoryear{{Schaye}, {Dalla Vecchia}, {Booth}, {Wiersma},
  {Theuns}, {Haas}, {Bertone}, {Duffy}, {McCarthy} \& {van de Voort}}{{Schaye}
  et~al.}{2010}]{Schaye}
{Schaye} J.,  {Dalla Vecchia} C.,  {Booth} C.~M.,  {Wiersma} R.~P.~C.,
  {Theuns} T.,  {Haas} M.~R.,  {Bertone} S.,  {Duffy} A.~R.,  {McCarthy} I.~G.,
     {van de Voort} F.,  2010, \mnras, 402, 1536

\bibitem[\protect\citeauthoryear{{Seljak} \& {Zaldarriaga}}{{Seljak} \&
  {Zaldarriaga}}{1996}]{Seljak}
{Seljak} U.,  {Zaldarriaga} M.,  1996, \apj, 469, 437

\bibitem[\protect\citeauthoryear{{Spergel}, {Bean}, {Dor{\'e}}, {Nolta},
  {Bennett}, {Dunkley}, {Hinshaw}, {Jarosik}, {Komatsu}, {Page} \& et
  al.}{{Spergel} et~al.}{2007}]{Spergel07}
{Spergel} D.~N.,  {Bean} R.,  {Dor{\'e}} O.,  {Nolta} M.~R.,  {Bennett} C.~L.,
  {Dunkley} J.,  {Hinshaw} G.,  {Jarosik} N.,  {Komatsu} E.,  {Page} L.,    et
  al. 2007, \apjs, 170, 377

\bibitem[\protect\citeauthoryear{{Spergel}, {Verde}, {Peiris}, {Komatsu},
  {Nolta}, {Bennett}, {Halpern}, {Hinshaw}, {Jarosik}, {Kogut} \& et
  al.}{{Spergel} et~al.}{2003}]{Spergel03}
{Spergel} D.~N.,  {Verde} L.,  {Peiris} H.~V.,  {Komatsu} E.,  {Nolta} M.~R.,
  {Bennett} C.~L.,  {Halpern} M.,  {Hinshaw} G.,  {Jarosik} N.,  {Kogut} A.,
  et al. 2003, \apjs, 148, 175

\bibitem[\protect\citeauthoryear{{Springel}}{{Springel}}{2005}]{Springel05}
{Springel} V.,  2005, \mnras, 364, 1105

\bibitem[\protect\citeauthoryear{{Springel}, {White} \& {Hernquist}}{{Springel}
  et~al.}{2001}]{Springel}
{Springel} V.,  {White} M.,    {Hernquist} L.,  2001, \apj, 549, 681

\bibitem[\protect\citeauthoryear{{Tasitsiomi}, {Kravtsov}, {Gottl{\"o}ber} \&
  {Klypin}}{{Tasitsiomi} et~al.}{2004}]{Tasitsiomi}
{Tasitsiomi} A.,  {Kravtsov} A.~V.,  {Gottl{\"o}ber} S.,    {Klypin} A.~A.,
  2004, \apj, 607, 125

\bibitem[\protect\citeauthoryear{{Teyssier}, {Pontzen}, {Dubois} \&
  {Read}}{{Teyssier} et~al.}{2013}]{Teyssier13}
{Teyssier} R.,  {Pontzen} A.,  {Dubois} Y.,    {Read} J.~I.,  2013, \mnras,
  429, 3068

\bibitem[\protect\citeauthoryear{{van den Bosch}}{{van den Bosch}}{2002}]{van}
{van den Bosch} F.~C.,  2002, \mnras, 331, 98

\bibitem[\protect\citeauthoryear{{van den Bosch}, {Jiang}, {Hearin},
  {Campbell}, {Watson} \& {Padmanabhan}}{{van den Bosch}
  et~al.}{2014}]{vandenBosch14}
{van den Bosch} F.~C.,  {Jiang} F.,  {Hearin} A.,  {Campbell} D.,  {Watson} D.,
     {Padmanabhan} N.,  2014, \mnras, 445, 1713 (vdB14)

\bibitem[\protect\citeauthoryear{{Wang} \& {White}}{{Wang} \&
  {White}}{2009}]{Wang09}
{Wang} J.,  {White} S.~D.~M.,  2009, \mnras, 396, 709

\bibitem[\protect\citeauthoryear{{Wechsler}, {Bullock}, {Primack}, {Kravtsov}
  \& {Dekel}}{{Wechsler} et~al.}{2002}]{Wechsler}
{Wechsler} R.~H.,  {Bullock} J.~S.,  {Primack} J.~R.,  {Kravtsov} A.~V.,
  {Dekel} A.,  2002, \apj, 568, 52

\bibitem[\protect\citeauthoryear{{White}}{{White}}{1996}]{White96}
{White} S.~D.~M.,  1996, in {Schaeffer} R.,  {Silk} J.,  {Spiro} M.,
  {Zinn-Justin} J.,  eds, Cosmology and Large Scale Structure {Formation and
  Evolution of Galaxies}.
p.~349

\bibitem[\protect\citeauthoryear{{Zel'dovich}}{{Zel'dovich}}{1970}]{ZelDovich}
{Zel'dovich} Y.~B.,  1970, \aap, 5, 84

\bibitem[\protect\citeauthoryear{{Zhao}, {Jing}, {Mo} \& {B{\"o}rner}}{{Zhao}
  et~al.}{2009}]{Zhao09}
{Zhao} D.~H.,  {Jing} Y.~P.,  {Mo} H.~J.,    {B{\"o}rner} G.,  2009, \apj, 707,
  354

\bibitem[\protect\citeauthoryear{{Zhao}, {Mo}, {Jing} \& {B{\"o}rner}}{{Zhao}
  et~al.}{2003}]{Zhao03}
{Zhao} D.~H.,  {Mo} H.~J.,  {Jing} Y.~P.,    {B{\"o}rner} G.,  2003, \mnras,
  339, 12

\end{thebibliography}
\bibliographystyle{mn2e}

\appendix
\section{Simulations and cosmology}\label{Simulations_appendix}

\begin{table*}
\centering  
\caption{List of simulations.  From
  left-to-right the columns show: simulation identifier; comoving box size;
  number of dark matter particles; dark matter
  particle mass; comoving (Plummer-equivalent) gravitational
  softening; maximum physical softening.} 
\label{sims}
\begin{tabular}{lrrrrr}
\hline
  Simulation & $L$ & $N$ & $m_{\rm{dm}}$ & $\epsilon_{\rm{com}}$ & $\epsilon_{\rm{prop}}$ \\ 
   & ($h^{-1}\rm{Mpc}$) & & ($h^{-1}\rm{M}_{\sun}$) & ($h^{-1}\rm{kpc}$) & ($h^{-1}\rm{kpc}$) \\  \hline\hline
   DMONL$Y_{-}$WMAP5$_{-}$L400N512 & 400 & $512^{3}$ & $3.4\times 10^{10}$ & 31.25 & 8.00\\
   DMONL$Y_{-}$WMAP5$_{-}$L200N512 & 200 & $512^{3}$ & $3.2\times 10^{9}$ & 15.62 & 4.00\\
   DMONL$Y_{-}$WMAP5$_{-}$L100N512 & 100 & $512^{3}$ & $5.3\times 10^{8}$ & 7.81 & 2.00\\
   DMONL$Y_{-}$WMAP5$_{-}$L050N512 & 50 & $512^{3}$ & $6.1\times 10^{7}$ & 3.91 & 1.00\\
   DMONL$Y_{-}$WMAP5$_{-}$L025N512 & 25 & $512^{3}$ & $8.3\times 10^{6}$ & 2.00 & 0.50\\ \hline
\end{tabular}
\end{table*}

\begin{table}
\centering  
\caption{Cosmological parameters.} 
\label{cosmo}
\begin{tabular}{lllllll}
\hline
  Simulation & $\Omega_{m}$ & $\Omega_{\Lambda}$ & $h$ & $\sigma_{8}$ & $n_{s}$ \\ \hline\hline
   DMONLY$_{-}$WMAP1 & 0.25 & 0.75 & 0.73 & 0.90 & 1.000\\
   DMONLY$_{-}$WMAP3 & 0.238 & 0.762 & 0.73 & 0.74 & 0.951\\
   DMONLY$_{-}$WMAP5 & 0.258 & 0.742 & 0.72 & 0.796 & 0.963\\
   DMONLY$_{-}$WMAP9 & 0.282 & 0.718 & 0.70 & 0.817 & 0.964\\
   DMONLY$_{-}$Planck1 & 0.317 & 0.683 & 0.67 & 0.834 & 0.962\\
\hline
\end{tabular}
\end{table}
  
In this work, we use the set of cosmological DMONLY simulations from the OWLS project (\citealt{Schaye}). The particle masses and gravitational softenings for each of these simulations are listed in Table \ref{sims}. Table \ref{cosmo} lists the sets of cosmological parameters adopted in the different simulations. In this section we describe the numerical techniques used to compute merger trees. 

The first step towards studying the mass assembly history of halos is to identify gravitationally bound structures and build halo merger trees. We begin by selecting the largest halo in each FoF group (\citealt{Davis}; we use a linking length of 0.2) (i.e. the main subhalo of FoF groups that is not embedded inside larger halos). Halo virial masses and radii are determined using a spherical overdensity routine within the SUBFIND algorithm (\citealt{Springel}) centred on the main subhalo of FoF halos. For a complete description on the method used to build the halo merger trees, see Paper II.

\section{Cosmology dependence}\label{cosmo_dependence}

The adopted cosmological parameters affect the halo MAH so that the larger $\sigma_{8}$ or $\Omega_{\rm{m}}$, the earlier halos assemble. As the formation time increases with increasing $\sigma_{8}$ or $\Omega_{\rm{m}}$, so does concentration. Therefore, in this section, we analyse how the change in cosmology affects our $c-M$ relation model. 

Our $c-M$ model described in Section \ref{cM_model} relies on the halo mass accretion history model, which we showed in Section \ref{MzLCDM} to be suitable for any cosmology. However, our semi-analytic model for halo concentration also relies on the formation redshift through the best-fitting relation given by eq. (\ref{rho_formation}),

$$\frac{\langle\rho\rangle(<r_{-2},z_{i})}{\rho_{\rm{crit}}(z_{i})} = A\times\frac{\rho_{\rm{crit}}(z_{-2})}{\rho_{\rm{crit}}(z_{i})},$$

\noindent where $A$ depends on cosmology ($A=887$ for WMAP5). 

We investigate the cosmology dependence of $A$ by following the analysis done in Section 4.1 and using the simulations with different cosmologies listed in Table \ref{cosmo}. We calculate the best-fitting $\langle\rho\rangle(<r_{-2},z_{i})-\rho_{\rm{crit}}(z_{-2},z_{i})$ relation to obtain the parameter $A_{\rm{cosmo}}$, where cosmo is WMAP1, WMAP3, WMAP9 or Planck. We found that if we keep $A=887$ fixed when computing the concentration-mass relations for cosmologies other than WMAP5, we obtain relations that are in very good agreement with the various relations in the literature. However, we obtain even better agreement with the simulations when we let $A$ vary slightly with cosmology. We found that $A_{\rm{WMAP1}} = 853$, $A_{\rm{WMAP3}} = 850$, $A_{\rm{WMAP9}} = 950$ and $A_{\rm{Planck}} = 880$, reproduce the $c-M$ relations best. For a calculation of $c-M$ relations in cosmologies other than the ones listed above, we recommend the reader to set $A=887$ fixed.

It is important to note that if the $A$ parameter is in fact cosmology dependent then halos which formed at different formation times (e.g. $z_{-2,\rm{cosmo1}}$ and $z_{-2,\rm{cosmo2}}$), but that correspond to the same background density ($\rho_{\rm{crit}}(z_{-2,\rm{cosmo1}})=\rho_{\rm{crit}}(z_{-2,\rm{cosmo2}})$), will have different concentrations ($c_{\rm{cosmo1}}\neq c_{\rm{cosmo2}}$). This implies that other factors affect the dark matter halo density profiles, and that the halo MAH alone is not sufficient for predicting concentrations.

\subsection{Fitting functions for the $c-M$ relation}

In this section, we provide fitting functions for the $c-M$ relation in the high-$z$ and low-$z$ regimes for Planck cosmology. The following expression is suitable for the low-redshift regime ($z\le 4$) and at all halo masses,

\begin{eqnarray}\nonumber
\log_{10} c &=& \alpha+\beta\log_{10}(M/\Msol)[1+\gamma(\log_{10}M/\Msol)^{2}],\\\nonumber
\alpha &=& 1.7543-0.2766(1+z)+0.02039(1+z)^{2},\\\nonumber
\beta &=& 0.2753+0.00351(1+z)-0.3038(1+z)^{0.0269},\\\nonumber
\gamma &=& -0.01537+0.02102(1+z)^{-0.1475}.
\end{eqnarray}

In the high-redshift regime the $c-M$ relation can be fitted using only two parameters. The following expression is suitable for $z>4$ and at all halo masses,

\begin{eqnarray}\nonumber
\log_{10} c &=& \alpha+\beta\log_{10}(M/\Msol),\\\nonumber
\alpha &=& 1.3081-0.1078(1+z)+0.00398(1+z)^{2},\\\nonumber
\beta &=& 0.0223-0.0944(1+z)^{-0.3907}.
\end{eqnarray}

\section{Online material}\label{script_details}

The concentration-mass relation model presented in this work, as well as the halo mass accretion history model, are included in the code named {\tt{COMMAH}} for COncentration-Mass relation and Mass Accretion History, available at \href{https://bitbucket.org/astroduff/commah}{\it{https://bitbucket.org/astroduff/commah}} and \href{http://astro.physics.unimelb.edu.au/Research/Public-Data-Releases/COMMAH}{\it{http://astro.physics.unimelb.edu.au/}} in Research/Public-Data-Releases/COMMAH. The code is available in both the python and IDL languages. Also, {\tt{COMMAH}} is in the PyPi python package, to install it type `pip install commah'. In this section we present a short overview of {\tt{COMMAH}}.

{\tt{COMMAH}} is a routine that follows the analytic model described in Section \ref{Halo_history_modelling} to calculate the MAH of a halo of mass $M_{0}$ at $z=0$ in any given redshift interval (e.g. $M(z)$ between $z=0$ and $10$). Also, {\tt{COMMAH}} calculates halo concentrations following the semi-analytic model described in Section \ref{cM_model}, and outputs the $c-M$ relation at any given redshift. In addition, it also computes the dark matter accretion rate, the rms of the density field, peak height, and the integral of the NFW density profile (see eq. \ref{integral}), suitable for DM annihilation calculations. {\tt{COMMAH}} calculates concentration solving eqns.~(\ref{zf_relation}) and (\ref{eq_c1}) by performing a Levenberg-Marquardt method. As described in the previous section, commah is suitable for any cosmology.

\end{document}